\documentclass[]{interact}

\usepackage{epstopdf}
\usepackage[caption=false]{subfig}

\usepackage{bbm}
\usepackage{algorithm}
\usepackage{multirow}
\usepackage{algpseudocode}
\usepackage{float}
\usepackage{multicol}
\usepackage{tabularx,booktabs,siunitx}
\usepackage{xcolor}
\usepackage[natbibapa,nodoi]{apacite}
\algdef{SE}[DOWHILE]{Do}{doWhile}{\algorithmicdo}[1]{\algorithmicwhile\ #1}%

\theoremstyle{plain}
\newtheorem{theorem}{Theorem}[section]
\newtheorem{lemma}[theorem]{Lemma}
\newtheorem{corollary}[theorem]{Corollary}
\newtheorem{proposition}[theorem]{Proposition}

\theoremstyle{definition}
\newtheorem{definition}[theorem]{Definition}
\newtheorem{example}[theorem]{Example}

\theoremstyle{remark}
\newtheorem{remark}{Remark}

\renewcommand*{~}{\relax\ifmmode\sim\else\nobreakspace{}\fi}

\newcommand{\Var}{\mbox{Var}}
\newcommand{\Varhat}{\widehat{\mbox{Var}}}

\newcommand{\fhat}{\hat{f}}

\newcommand{\rhohat}{\hat{\rho}}
\newcommand{\sigmahat}{\hat{\sigma}}

\newcommand{\sigmatilde}{\widetilde{\sigma}}

\newcommand{\mcB}{\mathcal{B}}

\newcommand{\mcI}{\mathcal{I}}

\newcommand{\mcN}{\mathcal{N}}
\newcommand{\mcO}{\mathcal{O}}

\newcommand{\mcS}{\mathcal{S}}

\newcommand{\mcX}{\mathcal{X}}

\newcommand{\mbE}{\mathbb{E}}

\newcommand{\mbI}{\mathbb{I}}

\newcommand{\mbP}{\mathbb{P}}

\begin{document}

\title{SIMULATION MODEL CALIBRATION WITH DYNAMIC STRATIFICATION AND ADAPTIVE SAMPLING}

 \author{
 \name{Pranav Jain\textsuperscript{a}, Sara Shashaani\textsuperscript{a}, and Eunshin Byon\textsuperscript{b}\thanks{CONTACT S. Shashaani. Email: sshasha2@ncsu.edu}}
 \affil{\textsuperscript{a} Edward P. Fitts Department of Industrial and System Engineering, North Carolina State University, Raleigh, NC 27695, USA \\ \textsuperscript{b} Department of Industrial and Operations Engineering, University of Michigan, Ann Arbor, MI 48109, USA}
 }

\maketitle

\begin{abstract}
    Calibrating simulation models that take large quantities of multi-dimensional data as input is a hard simulation optimization problem. Existing adaptive sampling strategies offer a methodological solution. However, they may not sufficiently reduce the computational cost for estimation and solution algorithm's progress within a limited budget due to extreme noise levels and heteroscedasticity of system responses. We propose integrating stratification with adaptive sampling for the purpose of efficiency in optimization. Stratification can exploit local dependence in the simulation inputs and outputs. Yet, the state-of-the-art does not provide a full capability to adaptively stratify the data as different solution alternatives are evaluated. We devise two procedures for data-driven calibration problems that involve a large dataset with multiple covariates to calibrate models within a fixed overall simulation budget. The first approach dynamically stratifies the input data using binary trees, while the second approach uses closed-form solutions based on linearity assumptions between the objective function and concomitant variables. We find that dynamical adjustment of stratification structure accelerates optimization and reduces run-to-run variability in generated solutions. Our case study for calibrating a wind power simulation model, widely used in the wind industry, using the proposed stratified adaptive sampling, shows better-calibrated parameters under a limited budget. \end{abstract}

\begin{keywords}
   data-driven calibration, heteroscedasticity, post-stratification, trust-region optimization
\end{keywords}

\section{INTRODUCTION}\label{sec:Intro}

Many simulation models, particularly those used to emulate real engineering systems, have physically unobservable parameters. This can be due to having a simulation model that has simplified the real system's dynamics (similar to low-fidelity models), or due to the environmental characteristics in which the real system operates (for example, the geographical and weather-related effects for a particular location and time), or due to the need for removing the initial transient state of the simulation to prepare the simulated data for analysis in time-dependent simulation outputs (for example, the warm-up period for a steady-state analysis). Particularly, in the first two examples, differences between simulated and real data are addressed during the important practice of calibration~\citep{sargent2010verification,schruben1980establishing}. 

Calibration of simulation  experiments with real-world observations is generally done through metamodeling approaches, typically with Bayesian models~\citep{kennedy2001bayesian,pousi2013}. However, these methods do not scale well with the size of the data or the number of the calibration parameters~\citep{Jeong2023IJDS,JEONG2024AE}. In the simulation literature, the so-called \emph{direct model calibration} is one that formulates the problem as a simulation optimization where the empirical loss is minimized by searching for the optimal calibration parameters over their feasible space~\citep{tolk2017advances}[Chapter 3]. Global search methods such as simulated annealing (SA) or random search (RS), as well as meta-heuristics such as the genetic algorithm or particle swarm optimization are popular for such problems~\citep{parra2018,guzman2009calibration,liu2007using,tahmasebi2012simulation} with the downside of lacking computational efficiency and convergence guarantees. Efficient stochastic optimization methods have proven effective for simulation model calibration using stochastic gradient~\citep{liu2022parameter} or line search methods~\citep{yuan2012calibration}. 

\subsection{Calibration as a Simulation Optimization} In traditional calibration of stochastic simulations, the discrepancy between the simulated and observed output values are computed while the inputs are simulated following an input probability model---a common practice in calibrating epidemiological models~\citep{cheng2023modelling}. But in many simulation experiments, the output data as well as real (not simulated) input data is used. We term these calibration problems \emph{data-driven calibration} given that the input itself is directly queried from a real dataset of both inputs and outputs of the real-world system. 

For example, in wind power generation models, a simulation model depends on unobservable parameters, such as the \emph{wake} parameter that describes the effect of wind decay in downstream turbines. The wake parameter $\theta$ can depend on a wind farm's location and other local characteristics.
Suppose that $((x_i, y_i): i = 1, 2, \ldots, n)$ is a collection of observed wind characteristic vectors $x_i \in \mathbb{R}^q$ (inputs), and observed generated wind power $y_i\in \mathbb{R}^p$ (outputs) from a particular wind farm over time, where $p$ is the number of turbines in the wind farm. The simulation model generates, as output, the predicted power---a vector-valued function $h(\theta,x_i)$ in $\mathbb{R}^p$. The calibration problem involves finding the wake parameter that best matches simulated and real outputs, i.e., 
\begin{equation}
    \min_{\theta\in\mathbb{R}} \sum_{i=1}^n \|h(\theta,x_i)-y_i\|^2_2.\label{eq:erm}
\end{equation}
The problem above is also known as the \emph{empirical risk minimization} (ERM). The simulation model, if accurately tuned, can be used to make decisions about the real system. 
However, if $n$ is large, then enumerating all $n$ data points to evaluate the performance under each $\theta$ will be very inefficient because each $h(\theta,x_i)$ evaluation requires running the simulation. In this case, one might resort to sampling, which renders ERM a stochastic problem. Using randomly selected small-scale data can alleviate the computational burden, so long as we can accurately tune the simulation model without utilizing all the observed data points. But when data is noisy, small samples result in inaccurate inference  about the calibration parameter. In many applications such as those involving reliability (e.g., in wind power generation),  using suboptimal calibration parameters to make decisions about the real system can cause high-risk consequences. 

\subsection{Contributions}
Viewing calibration as a simulation-optimization process, where the expected discrepancy between the model and real data is minimized, raises the question: can one reduce the algorithm complexity, that is, the overall simulation model runs to find a robust calibration parameter? Often, under a pre-specified computational budget, answering this question concerns allocating effort for (i) \emph{exploration}, (ii) \emph{exploitation}, and (iii) \emph{estimation} in each iteration of the optimization algorithm \citep{gao2015sequential,andradottir2009balanced}. Better exploration of the calibration parameter space and finding near-optimal solutions requires reducing the per-iteration budget, that is, the cost of estimation, as much as possible. We will formalize this discussion with a survey on \emph{adaptive sampling stochastic optimization} methods used towards this goal in Section \ref{sec:math-background}. But what is unique about a calibration problem to obtain cheaper estimates throughout the optimization process?

Ample input and output data in calibration contexts suggest a potential for \emph{stratified sampling}---a well-known variance reduction technique wherein the input domain is divided into multiple disjoint sub-regions, each of which with discerning distributional behavior of the outputs due to heteroscedasticity. The benefit of partitioning the data comes from the fact that the weighted average of conditional output variances is at most as large as the unconditional variance, that is, $\mbE_B[\text{Var}(A|B)]\leq\text{Var}(A)$ for any two random variables $A$ and $B$. All stratified sampling research seeks to maximize this reduction in variance. Meanwhile, when used within optimization, one has to account for estimating the performance of many alternatives for the calibration parameter. Since the distribution of simulation outputs changes under each $\theta$, it is reasonable to consider that the optimal partitioning for each $\theta$ should also vary. Hence, a priori partitioning of the input space (before starting the optimization algorithm) may be suboptimal. Therefore, the challenge of using stratified sampling within optimization can involve, in addition to choosing the sample size in each stratum, determining the best stratification structure for each $\theta$ evaluated during optimization. With existing research addressing the former~\citep{jain2023wake}, in this paper we focus on the latter and its integration within the optimization algorithm. We explore whether or not there is efficiency gains in choosing input sub-regions carefully for each iteration if it can be done at a low cost.

To that end, in a survey of stratified sampling techniques for estimation (Section~\ref{sec:SS}), we investigate the risks in allocating budget to each stratum due to poor estimates of conditional variance of outputs, which can be particularly concerning if the strata boundaries keep changing during optimization. We discuss that, for dynamic strata, \emph{post-stratified sampling} is a safe choice due to keeping the sampling distribution independent of the stratification structure. That is, we adopt independent sampling from the entire input space and then assign weights to each sample based on the placement of strata. We further leverage post-stratification variance as the metric when dynamically choosing the strata.

Next, we present two approaches to finding effective stratification structures (Section~\ref{sec:dynamic_strata}). The first method divides the data via binary trees (BT) for a greedy optimization of an objective function different from the standard trees. We propose an \emph{information gain} metric to evaluate the reduction in variance achieved after splitting a node and use that to choose the number of strata. The second method uses a linear relationship between some transformation of auxiliary or \emph{concomitant} variables (ConV) and the outputs to approximate closed-form boundaries. Sometimes, rapidly computable conditional means of input (not simulated) data as concomitant variables can be used, reducing the small simulated samples risks on strata misspecification. We extend the closed-form derivations for variables to simulated data and propose ideas to choose the best concomitant variable and the number of strata in each iteration. Increased learning ability by using all available data without needing new simulation runs and faster computation are the advantages of the second approach. Yet, it uses one concomitant variable at a time and hence may be less flexible than the first approach, which can stratify multiple variables jointly.

Lastly, we conduct a thorough numerical experiment (Section~\ref{sec:results}) with dynamically stratified adaptive sampling with BT and ConV and compare their performance with Bayesian optimization (BO), SA, and RS on Monte Carlo and discrete-event simulation toy examples. 
Further, implementation in a case study for the wind power simulation model calibration proves the effectiveness of the proposed methods in a real-world application and provides insights on their sensitivity and consistency. Comparisons are summarized with emphasis on remaining gaps and open questions for future research (Section~\ref{sec:conclusion}).

\section{MATHEMATICAL BACKGROUND AND CONTRIBUTIONS}
In this section, we define the data-driven calibration problem as a simulation optimization. We review its computational challenges, to remedy which we focus on a particular class of optimization algorithms that uses adaptive sampling within trust regions. We then list our contributions and gained insights that render this algorithm more successful for calibration.

\subsection{Problem Formulation and Standing Assumptions}
Consider the random instances of data $(X,Y)$ being generated from an underlying joint probability distribution and let $h(\theta,X)$ be the simulated random output corresponding to the vector pair $(X,Y)$. Then, defining the loss function $\ell(h(\theta,X),Y)$ as a measure of discrepancy between $h(\theta,X)$ and $Y$, the objective function becomes 
 \begin{align}
 \min_{\theta\in\Theta} \quad & f(\theta):= \mbE_{X,Y}[\ell(h(\theta,X),Y)]. 
            \label{eq:erms}
    \end{align}
The function $f(\theta)$ is non-negative, and we assume that it is nonconvex (due to nonconvexity of $h(\cdot,X)$) but continuously differentiable with Lipschitz continuous gradients, i.e., there exists a constant $L < \infty$ such that $\|\nabla_\theta f(\theta_1) - \nabla_\theta f(\theta_2)\|\leq L\|\theta_1-\theta_2\|$ for any $\theta_1,\theta_2\in\Theta$. For the remainder of this paper, we simplify the random objective function notation with $F(\theta,(X,Y)) := \ell(h(\theta,X),Y)$. The actual joint probability distribution in~\eqref{eq:erms} is unknown. Thus, we estimate the expected value at a particular $\theta$ in \eqref{eq:erms} via Sample Average Approximation (SAA) using a random sample set $\mcS$ of size $N$ from the available dataset, i.e., \[\fhat(\theta,N):= \frac{1}{N}\sum_{( x_j,y_j)\in\mcS} F(\theta, (x_j,y_j)).\] In other words, \emph{we consider each evaluation of the objective function for a single data point as one simulation replication}, and the total evaluations throughout the optimization are accounted for as the computational budget needed to reach an optimal solution. Under the Big Data context, i.e., large $n$, the selection of the points will be considered in an identically distributed and independent (i.i.d.) fashion. Most simulation models $h(\theta,\cdot)$ are too complex to have direct access to their derivatives with respect to $\theta$. Although direct gradients can be computable via techniques such as infinitesimal perturbation analysis~\citep{suri1987infinitesimal}, in most instances, that requires additional programming and analysis that may not be a feasible option in many applications. Hence, we consider the simulation model as a complete black-box and assume that $\nabla_\theta F(\theta,\cdot)$ are unavailable, which makes this problem a \emph{derivative-free optimization} (DFO)~\citep{conn2009introduction}. 

\subsection{Efficiency Challenges for Derivative-Free Simulation Optimization} DFO problems are much harder to solve due to the needed extra effort to approximate gradients through derivative-free methods. Therefore, the main challenge is: can we obtain good solutions with an optimization algorithm in this setting 
given a fixed computational budget? The answer to this question involves the trade-off between exploration and exploitation that, while primarily known in Bayesian optimization, is a general challenge with optimization algorithms evaluated in finite time. In a deterministic viewpoint, exploitation refers to evaluating the objective function value at multiple $\theta$'s within a sub-region to track it locally. Expending a lot of budget for exploitation leaves less budget for the algorithm to explore other regions of the search space. Using the objective function's structure to determine the number of $\theta$'s is not a viable option in DFO. Instead, DFO solvers expend budget to approximate the gradients with interpolation or finite differencing, among other methods. In the stochastic DFO, if the simulation outputs are very noisy, the approximated gradient can be inaccurate, and these methods can struggle to reach good solutions. Hence, exploitation in a stochastic setting involves both the number of $\theta$'s visited to approximate the gradient and estimating the objective function at each of those $\theta$'s.

Trust-region (TR) methods are increasingly known to be effective for non-convex DFO problems compared to line search or stochastic gradient methods due to their strict control of step size (tuned automatically throughout the search) and their implicit use of curvature by constructing a quadratic local model \citep{yuan2015recent}. The performance of TR depends on the quality of these local models as their high-quality can consistently identify good steps and progress per iteration.

However, the challenge with building high-quality models stems from the estimation error, which with $N$ Monte Carlo samples is inversely proportional to $\sqrt{N}$. Thus, given a fixed budget, having reliable estimates that can help build better local models (exploitation) comes at the cost of losing the budget to take more steps (exploration). An efficient algorithm appropriately determines the sample size at each point within the local sub-region while keeping enough budget for exploration. A successful strategy for this trade-off \emph{adapts} the choice of $\mcS(\theta)$ as a function of $\theta$ to the variability of $F(\theta,\cdot)$ and to the precision stipulated for convergence, i.e., more accurate models and function estimates as the algorithm nears the optimal region~\citep{byrd2012sample}. Recently, a TR-based algorithm that incorporates adaptive sampling for the DFO problems, and hence is appropriate for data-driven calibration, has been developed, called ASTRO-DF (Adaptive Sampling Trust-Region Optimization---Derivative-Free) \citep{shashaani2018astro,shashaani2016astro}. In this paper, we will use this algorithm as an instance of adaptive sampling solvers and investigate on how to tailor it for data-driven calibration. As discussed in Section~\ref{sec:Intro}, our goal will be to integrate dynamic stratification within this solver.

\subsection{Adaptive Sampling Trust-Region Optimization---Derivative-Free}

ASTRO-DF is an almost surely convergent simulation optimization solver for nonconvex problems that builds a local model in each iteration via interpolation and decides the number of simulation runs (samples) based on a proxy for optimality gap. The adaptive sample size guarantees the fastest proven sample complexity of $\mcO(\epsilon^{-4}
\log \epsilon^{-1})$ to reach $\epsilon$-optimality~\citep{ha2023}. Had the direct gradient observations been available, the lower-complexity solver (ASTRO---the derivative-based version of ASTRO-DF) could be used~\citep{vasquez2019}.

To better understand the sampling mechanism, let us briefly review the TR methods. For $\theta_k$ as the iterate (incumbent solution) at iteration $k$, a TR is defined as a closed ball around $\theta_k$, $\mcB_k = \{ \theta : \| \theta - \theta_k \|_2 \leq \Delta_k \}$, where $\Delta_k$ is the TR radius. A local model $M_k(\cdot)$ is then fitted to estimated objective function values at multiple $\theta$'s inside $\mcB_k$. This model suggests a candidate for the next incumbent, $\tilde{\theta}_{k+1}$ by predicting where the function will be minimized within $\mcB_k$. The reduction in the estimated objective function value at $\tilde{\theta}_{k+1}$ is then compared to the corresponding reduction in the model. If sufficient reduction in the objective function value is achieved at $\tilde{\theta}_{k+1}$, then the candidate solution is accepted, i.e., $\theta_{k+1} = \tilde{\theta}_{k+1}$, and the TR expands. If rejected, then $\theta_{k+1} = \theta_k$, the TR radius shrinks, and a new model is constructed in a smaller neighborhood around $\theta_k$. For a complete listing of the new variant of this algorithm that uses the new proposed approaches for dynamic stratification, see Section~\ref{sec:alg}. 

In a (deterministic) DFO setting, the TR model gradient (in Euclidean norm) is maintained in lock-step with the TR radius, i.e., $\|\nabla M_k(\theta_k)\|=\mcO(\Delta_k)$~\citep{conn2009introduction}. Then, proving that $\lim_{k \to \infty} \Delta_k = 0$ guarantees that the model gradient converges to 0. Handling the stochasticity comes in when one also needs to maintain a lock-step between the model gradient and the true function gradient. This is, in effect, what a high-quality model needs to accomplish in every iteration. ASTRO-DF deals with this challenge by choosing the optimal (i.e., most efficient) sample size at each visited $\theta$. Since the sequence $\{\Delta_k\}$ shrinks as the algorithm nears optimality, ASTRO-DF uses the fourth power (appropriately selected to maintain model quality) of the TR radius as the acceptable upper bound for the standard error at each visited $\theta$. The sample size is therefore a \emph{stopping time} of the form 
\begin{equation}
    N_k=\min\left\{n\geq \lambda_k: \sqrt{\Varhat(\fhat(\theta_k,n))}\leq \kappa\frac{\Delta_k^2}{\sqrt{\lambda_k}}\right\},\label{eq: adaptive_sampling}
\end{equation} since with every added sample, the LHS (standard error estimate) changes and eventually reduces while the RHS (slightly deflated optimality gap proxy) remains unchanged. 

In~\eqref{eq: adaptive_sampling}, $\lambda_k$ is a deterministic sequences that increases logarithmically with $k$, and $\kappa$ is a positive constant. The deflation ensures that the acceptable standard error threshold is stricter in the later iterations and is essential for proving almost sure model quality guarantees (and hence algorithm convergence)~\citep{ha2023adaptive}. Another role of $\lambda_k$ is lower bounding the sample size so that even under early stopping due to a poor estimate of the standard error, $N_k$ increases at least logarithmically to increase estimation accuracy. The algorithm first runs $\lambda_k$ i.i.d. replications to obtain $\Varhat(\fhat(\theta_k,\lambda_k))$ and, if needed, adds one sample at a time. As a result, the adaptive sample size is small during initial iterations when the optimality gap is large and increases in the later iterations when the algorithm appears to have neared optimality. 

Choosing $N_k$ provides theoretical guarantees for efficiency, but since there is no upper limit to the stopping time, it can practically be very large due to high noise in $F(\theta_k,\cdot)$. High level of  noise is notoriously present in data-driven calibration causing extremely large sample sizes, which is undesirable under a finite budget setting. Enhancing the algorithm with a variance reduction technique such as stratified sampling (leveraging conditional behavior of the outputs in input sub-regions) can help avoid such larger sample sizes. However, a seamless incorporation of stratified sampling with adaptive sampling is challenging as which stratum to sample from in each recursion of the stopping time induces more uncertainty to the algorithm. There are also risks and opportunities in selecting the strata themselves appropriately in each iteration.

\section{STRATIFIED SAMPLING FOR OPTIMIZATION} \label{sec:SS}
Stratified sampling groups similar data into strata such that the output within each group is similar and between any two groups is different. This helps learn the heterogeneity in the data for estimation, which leads to variance reduction \citep{ROSS2013153}. Intuitively, to efficiently allocate overall samples to each stratum, more points should be sampled from a stratum with higher variance. This efficient allocation of the computational budget can reduce the variance of the estimators and expedite the optimization. The impact of stratified sampling on the optimization routine is influenced by (i) the allocation scheme used to determine the sample size of each stratum and (ii) the stratification structure.

\paragraph*{The allocation scheme} depends on what sampling strategy is utilized. \emph{Proportional} allocation sets the sample size of a stratum based on the probability of picking a point from that stratum. \emph{Optimal} allocation depends on the above probability and the output variance in that stratum \citep{neyman1934allocation}. An inaccurate estimate of these two values can reduce the effectiveness of stratified sampling and produce worse estimates of the performance. Thus, many studies in the literature have averted to explore different ways to solve the problem of sample size allocation. Mathematical techniques like convex programming \citep{huddleston1970optimal}, branch and bound methods \citep{bretthauer1999nonlinear}, and delta method \citep{glynn2021efficient} have been proposed to determine optimal allocation. Since optimal allocation can be erratic if the variance cannot be estimated accurately, a hybrid allocation scheme that switches between proportional and optimal allocation as more insights are gained by running simulation can also be used \citep{pettersson2021adaptive}. Another common method is an adaptive optimal allocation that minimizes the variance within each stratum \citep{etore2010adaptive,kawai2010asymptotically}. All these studies focus on applying stratified sampling for simulations or statistical inference. Within the optimization framework, the optimal allocation has been implemented via batching \citep{chen2018novel,hassan2006non,zhao2014accelerating}. One of the drawbacks of batching is that the sample sizes can be larger than specified by adaptive sampling and may result in inefficient budget utilization.

\paragraph*{The stratification structure} can be determined by  partitioning the data based on an input variable such that output data in each stratum exhibits similar probabilistic characteristics. Consequently, we can assume a separate conditional distribution for the outputs in each stratum. This problem has been widely analyzed for one-time stratification (for inference) using heuristics like clustering \citep{farias2020similarity, tipton2013stratified, zhao2014accelerating}, genetic algorithms \citep{keskinturk2007genetic}, binary trees \citep{jain2021wake,jain2022robust}, etc. A drawback of these heuristics and greedy search methods is their reliance on the data used to build the structure, being susceptible to poor performance if the data is noisy or insufficient. Another approach is to use a theoretically derived closed-form solution to divide the data via concomitant variables \citep{dalenius1950problem}. Concomitant variables are traditionally simulated input data with known mean and variance. If the concomitant variables are correlated to the outputs, the strata boundaries that minimize the variance can be determined by solving implicit equations derived given their conditional distributions  \citep{dalenius1951problem, taga1967optimum}. However, the closed-form boundaries' equations are only solvable if the concomitant variables follow a well-known probability distribution~\citep{sethi1963note,singh1969optimum}. Otherwise, iterative fixed-point methods~\citep{cochran1977sampling, sethi1963note}, convex optimization~\citep{brito2010exact,de2017optimization}, and dynamic programming~\citep{khan2008determining} have been used to solve these equations. Using these approaches for optimization can be computationally expensive given that they will be invoked at every iteration of the algorithm. In addition, they are only applicable when the number of strata and the stratification variable are known a priori, both of which may also vary from one iterate to another during optimization.

\subsection{Notations and Definitions}
Consider a stratification structure $\mcI_k$, which divides the input space into $Z_k$ disjoint sets ($\mcX_{k,1},\mcX_{k,2},\cdots,\mcX_{k,Z_k}$) such that $\mcX := \cup_{z=1,\ldots,Z_k} \mcX_{k,z}$ is the whole input space. 
 For a sample $\mcS_k:=\mcS(\theta_k)$ of size $N_k$ formed by subsamples $\mcS_{k,z}=\{(x_j,y_j)\in\mcS_k:\ x_j \in \mcX_{k,z}\}$ of size $N_{k,z}$ in each stratum (i.e., $N_k = \sum_{z=1}^{Z_k} N_{k,z}$), the estimated stratified sampling mean is 
\begin{equation}
    \fhat_{\text{strat}}(\theta_k,N_k|\mcI_k) = \sum_{z=1}^{Z_k} p_{k,z} \fhat_{z}(\theta_k),\label{eq:st_mean}
\end{equation}
where $p_{k,z} = \mbE[\mbI{1}{(X\in \mcX_{k,z})}]$ is the probability of drawing a sample whose input lies in stratum $z$, and $\fhat_{z}(\theta_k)$ is the sample average in stratum $z$, i.e., 
\begin{equation*}
    \fhat_{z}(\theta_k) = \frac{1}{N_{k,z}} \sum_{(x_j,y_j)\in\mcS_{k,z}} F(\theta_k, (x_j,y_j)).
\end{equation*}Throughout this article, we assume $p_{k,z} = \frac{|\mcX_{k,z}|}{|\mcX|}$ given the big data setting. The variance of the stratified sampling estimator is 
\begin{equation*}
    \text{Var}(\fhat_{\text{strat}}(\theta_k,N_k|\mcI_k)) = \sum_{z=1}^{Z_k} \frac{p_{k,z}^2 \sigma^2_{k,z}}{N_{k,z}},
\end{equation*}
where $\sigma^2_{k,z} := \mbE\left[\left(F(\theta_k,(X,Y)) - f_{z}(\theta_k)\right)^2\middle\vert X \in \mcX_{k,z}\right]$ is the variance of outputs in stratum $z$ with $f_{z}(\theta_k):=\mbE\left[F(\theta_k,(X,Y))\middle\vert X\in \mcX_{k,z}\right]$ as its mean. The estimated mean and variance of the stratified sampling estimator depend on the stratification structure $\mcI_k$ and the set of sampled points $\mcS_k$. The reduction in the variance of the stratified sampling estimator given $\mcI_k$ depends on how the samples are allocated to each stratum.

\subsection{Review: Proportional vs. Optimal Allocation}\label{sec: prop_vs_opt}
For ease of exposure, let $N_k$ be a deterministically growing sample size instead of a stopping-time sample size chosen adaptively for the remainder of this section. 
In proportional allocation, $N_{k,z}=p_{k,z}N_k$ whereas in optimal (or Neyman) allocation $N_{k,z}=w_{k,z}N_k$, with weights computed as  $$w_{k,z} = \frac{p_{k,z}\sigma_{k,z}}{\sum_{z'=1}^{Z_k} p_{k,z'}\sigma_{k,z'}}.$$ Optimal allocation results in the lowest variance if $\sigma_{k,z}$'s are known for all $z$ with $$\text{Var}(\fhat_{\text{os}}(\theta_k,N_k\mid \mcI_k))=\frac{1}{N_k}\left(\sum_{z=1}^{Z_k}p_{k,z}\sigma_{k,z}\right)^2\ \text{and Var}(\fhat_{\text{ps}}(\theta_k,N_k\mid \mcI_k))=\frac{1}{N_k}\sum_{z=1}^{Z_k}p_{k,z}\sigma_{k,z}^2,$$ for the optimal and proportional estimator, respectively. However, since estimates of the conditional variance $\sigma_{k,z}$ in each stratum, i.e., \begin{equation}
    \sigmahat_{k,z}^2:=\frac{1}{N_{k,z}-1}\sum_{(x_j,y_j)\in\mcS_{k,z}}(F(\theta_k,(x_j,y_j)) - \fhat_{z}(\theta_k))^2,\label{eq:var-est}
\end{equation} ought to be used instead, optimal allocation is subject to risks due to inaccurate $\sigmahat_{k,z}$'s, which is more prominent in the early iterations. On the other hand, when using proportional allocation, the sample size of stratum $z$ depends only on $p_{k,z}$, which can be more rapidly estimated using all the available input data. 
The maximum reduction in variance with optimal allocation is mostly effective in the later iterations for another reason too. Let $N'_{k,z}$ be the theoretical optimal sample size of stratum $z$, $\text{Var}_{\text{opt}}$ be the variance of optimal allocation, $\hat{N}_{k,z}$ be the estimated sample size of stratum $z$ with optimal allocation and $\text{Var}_\text{est}$ be the corresponding variance without stratification such that $N_k = \sum_z N'_{k,z} = \sum_z \hat{N}_{k,z}$. Increased variance due to under- or over-estimating the optimal allocation sample sizes can be characterized as
\begin{equation*}
    \text{Var}_\text{est} \geq \text{Var}_\text{opt}\left( 1 + \frac{1}{N_k} \sum_{z=1}^{Z_k} \frac{(\hat{N}_{k,z} - N'_{k,z})^2}{\hat{N}_{k,z}}\right),
\end{equation*}
which is more significant when $N_k$ is small~\citep{cochran1977sampling}. For early iterations of optimization with small $N_k$ (where the algorithm trajectory is most vulnerable for progress), using proportional allocation is almost as good as the reduction with optimal allocation~\citep{pettersson2021adaptive}. In fact, one can be more confident to obtain \emph{any} reduction in the variance with proportional allocation than with the optimal allocation~\citep{asmussen2007stochastic}[p. 151]. 
In summary, using proportional allocation for the purpose of optimization reduces the run-to-run variability of the algorithm. 

\subsection{Allocation Schemes with Adaptive Sampling}\label{sec:allocation}
Section~\ref{sec: prop_vs_opt} discusses batch-based allocation. However, in implementation of stratified sampling in combination with adaptive sampling rules such as those in ASTRO-DF, we need to allocate each sample sequentially, one at a time. First, \eqref{eq: adaptive_sampling} is replaced by the standard error on the LHS with the standard error of the stratified sampling estimator. Next, if the standard error exceeds the optimality gap, it is necessary to determine from which stratum should an additional point be sampled. A selective randomized method proposes selecting a random stratum to sample from using  
the probability mass function
\begin{equation}
\pi_{k,z}(n)=\Pr\{\text{selecting stratum }z\text{ after $n$ samples}\}=\frac{v_{k,z}(n)\mbI{1}\{v_{k,z}(n) > \frac{n_z}{n}\}}{\sum_{z'=1}^{Z_k} v_{k,z'}(n)\mbI{1}\{{v_{k,z'}(n)>\frac{n_{z'}}{n}\}}},
\label{eq:pmf_allocation}
\end{equation}
where $n_z$ is the current number of $n$ samples that belong to stratum $z$ and 
\[
v_{k,z}(n) = 
\begin{cases}
    \hat{w}_{k,z}(n), & \text{for optimal allocation} \\ p_{k,z}, & \text{for proportional allocation},
\end{cases}
\] with $\hat{w}_{k,z}$ as the estimate of $w_{k,z}$ using $n_z$ samples. 
The expected sample size in stratum $z$ for a fixed iterate $\theta_k$  conditional on the stopping time is 
\begin{align*}
     \mbE[N_{k,z}|N_k=n_k] & = \mbE\left[z \text{ selections in pilot runs} + \sum_{n=\lambda_k +1}^{n_k} 1\times \mbP\{z\text{ selected on }n\text{th run}\}\right]\\
     &\approx \mbE[v_{k,z}(\lambda_k)]\lambda_k + \sum_{n=\lambda_k +1}^{n_k}\mbE[\pi_{k,z}(n)].
 \end{align*}

The approximation reveals the complication with analyzing the sample size of each stratum. Although $v_{k,z}$ has less variability in proportional allocation than in optimal allocation, even proportional allocation can result in instability simply because $(N_{k,1},N_{k,2},\ldots,N_{k,Z_k})$ is a multinomial random vector with each mode's probability changing sequentially with every new sample added following \eqref{eq:pmf_allocation}. In summary, when using adaptive sampling, both allocation schemes are subject to the changing sampling distribution with every added sample. It can cause unstable updating of $\theta_k$ in the optimization process.

\subsection{Post-stratification for Stratified Adaptive Sampling with Changing Strata in Optimization}\label{sec:alg}
Stratified adaptive sampling has been explored using optimal allocation and its extensions for stochastic gradient methods \citep{espath2021equivalence,liu2022parameter} and Nelder Mead \citep{aguiar2022new}. \citep{jain2021wake,jain2022robust} examined how robust is the implementation of stratified adapting sampling for TR methods. However, stratified sampling requires stratification structure to be known a priori to sample points from each stratum independently. 
Consequently, most existing studies use a constant stratification structure, i.e., $\mcI_k=\mcI$ for all $k$, to maintain a consistent sampling framework throughout the search. Even with the fixed structure, stratified sampling with optimal (or proportional) allocation faces increased stochasticity when the sampling distribution changes in the optimization process, as discussed in Section~\ref{sec:allocation}. 
To overcome these vulnerabilities, one can use post-stratification, which first samples randomly from the entire 
population. Then the estimation follows similar to stratified sampling with proportional allocation using $N_{k,z}$, the number of sampled points that are within each stratum. Central limit theorems for proportional allocation apply to post-stratification~\citep{asmussen2007stochastic}, and its finite-time performance in queuing simulations has been on par with variance reduction obtained using control variates~\citep{wilson1984variance}. 

The post-stratified sampling estimator $\fhat_\text{post}(\theta_k,N_k|\mcI_k)$ is evaluated via \eqref{eq:st_mean} to obtain each $\fhat_{z}(\theta_k)$; its variance is then exactly computed~\citep{cochran1977sampling} as 
\begin{equation}
    \text{Var}(\fhat_\text{post}(\theta_k,N_k|\mcI_k))=\frac{1}{N_{k}}\sum_{z=1}^{Z_k} p_{k,z} \sigma^{2}_{k,z} + \frac{1}{N_{k}^2} \sum_{z=1}^{Z_k} (1 - p_{k,z}) \sigma^{2}_{k,z} + \mcO\left(\frac{1}{N_k^3}\right). \label{eq:post_var}
\end{equation}
The first term in \eqref{eq:post_var} is the variance of the proportional allocation, and the second term is the increase in variance because the post-stratification does not account for the stratification structure. Post-stratification is not an allocation scheme as the allocation happens automatically. Importantly, this reduces variability  since $(N_{k,1},N_{k,2},\ldots,N_{k,Z_k})$ is now a multinomial random vector with each mode's probability determined by $p_{k,z}$, and hence fixed. From \eqref{eq:var-est}, the reduced  variability manifests in more stable estimates for the conditional variance in each stratum. In other words, under adaptive sample sizes, $\sigmahat_{k,z}$ obtained via post-stratification is a better estimate for $\sigma_{k,z}$ than that obtained via the proportional allocation.

When we want to let the stratification structure $\mcI_k$ change with the decision variable $\theta_k$, post-stratification will again be more appropriate since it does not necessitate a priori knowledge of the stratification structure when drawing samples. Therefore, we can construct the stratification structure using the pilot simulations and estimating the post-stratified variance estimator \eqref{eq:post_var}. This means $\lambda_k$ in~\eqref{eq: adaptive_sampling} needs to be significantly larger than in the standard ASTRO-DF to start, but ultimately, it can save budget for exploration later in the search. In Section~\ref{sec:dynamic_strata}, we will present two approaches for constructing the strata from the learning yielded by the $\lambda_k$ pilot samples.  

Once the stratification structure is ready, we let the automatic allocation of samples to each stratum be executed while we only take i.i.d. samples from the input space and leverage the adaptive sampling (to decide when to stop taking more samples) with less effort. If the standard error of the post-stratified estimator is more than the RHS in~\eqref{eq: adaptive_sampling}, one point is randomly sampled from the entire data that will lie in one of the strata depending on the stratification structure. Then, the estimates are updated, and the adaptive sampling rule is examined again, and this process repeats until~\eqref{eq: adaptive_sampling} is satisfied. 
Algorithm~\ref{alg:astrodf-ps} outlines the implementation of ASTRO-DF for a given stratification structure $\mcI_k$, and the details of post-stratified adaptive sampling are summarized in Algorithm~\ref{alg:pas}. The output of Algorithm~\ref{alg:pas} will be denoted depending of whether the input is $\theta_k$ (iterate), $\theta_k^i$ (interpolation points), or $\tilde{\theta}_{k+1}$ (candidate solution).

\begin{algorithm}
\caption{ASTRO-DF with  Dynamic Post-Stratification}
\label{alg:astrodf-ps}
\begin{algorithmic}[1]
\item \textbf{Input:} Initial solution $\theta_0$ and TR radius $\Delta_0$, maximum budget $T$, and success threshold $\eta > 0$. 
\item \textbf{initialization:} Set $\text{calls} = 0$ and iteration $k = 0$. 
\While{$\text{calls} < T$}
 \State Estimate $\fhat_\text{post}(\theta_k,N_k|\mcI_k)$ and $\widehat{\text{Var}}(\fhat_\text{post}(\theta_k,N_k|\mcI_k))$ via Alg.~\ref{alg:pas}.
 \State Set $\text{calls} = \text{calls} + N_k$.
 \State Select $2d$ points using the coordinate basis in $\mcB_k$, i.e., $\{\theta_k\pm\Delta_k e_i\}_{i=1}^d$.
 \State Estimate interpolation points' function value with $N_k^i$ samples via Alg.~\ref{alg:pas}. 
 \State Set $\text{calls}=\text{calls}+ \sum_{i=1}^{2d}N_k^i$. 
 \State Construct model $M_k(\theta)$ by interpolation and find $\tilde{\theta}_{k+1}$, its minimizer in $\mcB_k$.
 \State Estimate $\fhat_\text{post}(\tilde{\theta}_{k+1},\tilde{N}_{k+1}|\tilde{\mcI}_{k+1})$ and $\widehat{\text{Var}}(\fhat_\text{post}(\tilde{\theta}_{k+1},\tilde{N}_{k+1}|\tilde{\mcI}_{k+1}))$ via Alg.~\ref{alg:pas}.
 \State Set $\text{calls}=\text{calls}+ \tilde{N}_{k+1}$.
 \State Compute the success ratio $\rhohat_k = \frac{\fhat_\text{post}(\theta_k,N_k|\mcI_k) - \fhat_\text{post}(\tilde{\theta}_{k+1},\tilde{N}_{k+1}|\tilde{\mcI}_{k+1})}{M_k({\theta}_k) - M_k(\tilde{\theta}_{k+1})}$.
 \If{$\rhohat_k > \eta$}
 \State Set $\theta_{k+1} = \tilde{\theta}_{k+1}$ and $\Delta_{k+1}>\Delta_k$.
 \Else
 \State Set $\theta_{k+1} = \theta_k$ and $\Delta_{k+1}<\Delta_k$.
 \EndIf
 \State Set $ k = k + 1$.
\EndWhile
\item \textbf{output:} Terminal solution $\theta_k$ and terminal performance $f(\theta_k)\approx\widehat{\mbE}[F(\theta_k,(X,Y))]$.
\end{algorithmic}
\end{algorithm}

\begin{algorithm}
\caption{Post-Stratified Adaptive Sampling}
\label{alg:pas}
\begin{algorithmic}[1]
\item \textbf{Input:} Available dataset $\mcX$, iterate $\theta_k$,  
TR radius $\Delta_k$, minimum sample size $\lambda_0$, and constant $\kappa>0$. 
 \State Compute $\lambda_k = \lceil \lambda_0 (\log k)^{1.5} \rceil$ 
 \State Run $\lambda_k$ i.i.d. simulations to obtain $F(\theta_k,(x_j,y_j)) \  \forall j=1,2,\dots,\lambda_k$. Set $N_k = \lambda_k$.
 \State Generate a stratification structure $\mcI_k$ with $Z_k$ strata via Alg.~\ref{alg:synch_cart} or Alg.~\ref{alg:cv-boundaries}.
 \State Compute $\fhat_\text{post}(\theta_k,N_k|\mcI_k),\widehat{\text{Var}}(\fhat_\text{post}(\theta_k,N_k|\mcI_k))$ with $N_k=\sum_{z=1}^{Z_k} N_{k,z}$ using $\mcI_k$.
 \While{$\sqrt{\widehat{\text{Var}}(\fhat_\text{post}(\theta_k,N_k|\mcI_k))} > \kappa\frac{\Delta_k^2}{\sqrt{\lambda_k}}$}
 \State Take an i.i.d. sample and identify stratum $z$ that it belongs to based on $\mcI_k$.
 \State Set $N_{k,z} = N_{k,z} + 1$ and $N_k = N_k + 1$.
 \State Update $\fhat_{z}(\theta_k), \fhat_\text{post}(\theta_k,N_k|\mcI_k),$ and $\widehat{\text{Var}}(\fhat_\text{post}(\theta_k,N_k|\mcI_k))$. 
 \EndWhile
\item \textbf{output:} Sample size $N_k$ and estimates  $\fhat_\text{post}(\theta_k,N_k|\mcI_k),\widehat{\text{Var}}(\fhat_\text{post}(\theta_k,N_k|\mcI_k))$.
\end{algorithmic}
\end{algorithm}

\section{DYNAMIC CONSTRUCTION OF STRATA}\label{sec:dynamic_strata}
An important aspect of Algorithm~\ref{alg:astrodf-ps} is determining the stratification structure $\mcI_k$. Constructing a stratification structure involves determining three things: 
\begin{itemize}
    \item[(i)] number of strata,
    \item[(ii)] stratification variable (when there are multiple input variables), and
    \item[(iii)] strata boundaries (split values).
\end{itemize}
A fixed structure can be built if physics-based (i)-(iii) are known a priori. In many practices, these will not be known, and while asynchronous or static stratification still has an advantage for variance reduction, not getting (i)-(iii) right in implementation may barely benefit the optimization if not slowing it down~\citep{jain2022robust}. The question is, can we do better than selecting the strata without consideration for the local conditional behavior of the objective function? Especially for heteroscedastic problems, fixed strata may not be optimal at every iteration as the conditional output distribution can greatly vary at different $\theta$'s. Synchronous or dynamic stratification may thus be beneficial in building the optimal strata for each $\theta$ that is evaluated during optimization if the computational cost of doing so is not too expensive.

Several methods have been proposed to build a dynamic structure via greedy search~\citep{etore2011adaptive, pettersson2021adaptive, liu2022parameter} that address (ii) and (iii) but assume there is always a fixed number of strata across iterations. More strata means more quantities that need to be estimated (mean and variance of each stratum). Obtaining maximal reduction in variance requires large samples in each stratum to accurately estimate their statistics. For too many strata, the budget utilization can thus be extremely high. 

In this work, we propose two ways to determine the stratification structure by finding solutions to (i)-(iii) simultaneously such that 
\begin{equation*}
    \mcI_k =  \underset{\begin{subarray}{c} \mcI \\ \text{Strata Structure}
    \end{subarray}}{\arg\!\min} {\widehat{\text{Var}}\left(\fhat_\text{post}(\theta_k,N_k=\lambda_k|\mcI)\right)}.
\end{equation*}
We propose two methods for stratification: a greedy search with a new variant of binary trees enabling more complex strata, and a closed-form solution with only one stratification variable at a time but with applicable to either \emph{real}  or \emph{simulated} data.

\subsection{Stratification via Binary Trees}

The first stratification method we present greedily divides the data with binary trees to minimize the estimated variance. The first step is to decide the stratification variable and the corresponding split point. Let $z$ be the current leaf that is to be split and $X^1, X^2, \ldots, X^p$ be the possible options for stratification variable with $\text{Rng}_{k,z}(X^t)$ as the set of all possible values that the $t$-th variable can take in leaf $z$. Splitting value will divide leaf $z$ into two candidate leaves 
defined for sets of (all not samples of) input data $\mcX_{k,z,l(t,x)}=\{X\in\mcX_{k,z}:\ X^t\leq x\}$ and $\mcX_{k,z,r(t,x)}=\{X\in\mcX_{k,z}:\ X^t> x\}$  to the left and right of the splitting criteria, respectively. We denote the sample size and the estimated variance in the left and right candidate leaf after splitting as $N_{k,z,l(t,x)},N_{k,z,r(t,x)}, \sigmahat^2_{k,z,l(t,x)}$, and $\sigmahat^2_{k,z,r(t,x)}$, respectively. Then, the optimal stratification variable and the corresponding split point are determined by minimizing the variance of proportional allocation estimated after splitting, i.e.,
\begin{equation}
\begin{aligned}
    (X^*_{k,z},x^*_{k,z}) = \underset{\begin{subarray}{c}
         t=1,2,\ldots,p  \\
        x \in \text{Rng}_{k,z}(X^t)
    \end{subarray}}{\arg\!\min} \,\,\,  &  \sigmahat^2_{k,z,l(t,x)} Q_{k,z,l(t,x)} +  \sigmahat^2_{k,z,r(t,x)} Q_{k,z,r(t,x)}. \label{eq:CART_Obj}
\end{aligned}
\end{equation} In~\eqref{eq:CART_Obj}, weights $Q_{k,z,l(t,x)}=\frac{N_{k,z,l(t,x)}}{N_{k}}=\frac{|\mcS_{k,z,l(t,x)}|}{|\mcS_{k}|}$ and $Q_{k,z,r(t,x)}=\frac{N_{k,z,r(t,x)}}{N_{k}}=\frac{|\mcS_{k,z,r(t,x)}|}{|\mcS_{k}|}$ are used from samples collected so far (i.e., with $|\mcS_k|=\lambda_k$ pilot samples) in place of probabilities $p_{k,z,l(t,x)}=\frac{|\mcX_{k,z,l(t,x)}|}{|\mcX|}$ and $p_{k,z,r(t,x)}=\frac{|\mcX_{k,z,r(t,x)}|}{|\mcX|}$ because finding the probabilities using the entire available data for every possible $x \in \text{Rng}_{k,z}(X^t)$ can be computationally expensive.

We denote the optimal left and right splits from solving~\eqref{eq:CART_Obj} with $l^*$ and $r^*$. This optimization can be solved sequentially as the binary tree grows to provide an ultimate stratification via greedy search. The optimal stratification variable at each leaf can be different, and it depends on the current iterate $\theta_k$ and the set of points sampled during that iteration. The split variables and values $X_{k,z}^*,x_{k,z}^*$ determine strata boundaries, which can change with $\theta_k$ and the sample size. Upon accepting a split, we update the indexing of the strata (leaves) by setting the index of the left candidate leaf as $z$ and the index of the right candidate leaf as $Z_k+1$ and finally incrementing the total strata by 1, i.e., $Z_k=Z_k+1$. This means the old $z$-th leaf is now replaced with the left candidate leaf, and the right candidate leaf is added to the list of leaves and will not be considered for further splitting.

The next step is to determine when the algorithm should stop splitting the data or, in other words, the number of strata; each leaf of the tree will be a stratum at the end. One approach is to pre-select the maximum number of strata. Too many strata means many quantities to be estimated, and too few strata can mean we lose substantial variance reduction. The best choice for $Z_k$ can differ from one iterate to another, and pre-selecting it is subjective. A natural solution can be cross-validation, whereby the prediction error that falls below a certain threshold would stop the stratification process. The issues with this approach are cutting the small sample of data used for constructing the tree even shorter to form cross-validation folds in addition to ad-hoc choice of the prediction error threshold, which could rely on problem-dependent hyperparameters. These issues can result in a suboptimal stratification structure, affecting the estimates and the optimization process. Instead, we present an approach that determines the necessity of splitting a leaf by assessing the extent of variance reduction achieved through the proposed split based on \emph{information gain} \citep{quinlan1986induction}. 

Consider $\mcI_k$ as the stratification structure built so far. Before splitting, we will collect a statistic from each leaf $z=1,2,\cdots,Z_k$. 
Let $\sigmahat^2_{k} = \widehat{\text{Var}}(\fhat_\text{post}(\theta_k,N_k|\mcI_k)$ be the estimated variance of the post-stratified estimator given the current structure $\mcI_k$, and the new variance of the post-stratified estimator if this leaf was selected for splitting using the criteria returned by~\eqref{eq:CART_Obj} be evaluated as
\begin{align}
    \sigmatilde_{k}^2(z) =\sigmahat_{k}^2 - \frac{1}{N_k}&\left((p_{k,z}+\frac{1-p_{k,z}}{N_k} )\sigmahat_{k,z}^2\right.\nonumber\\ 
    &\left.-(p_{k,z,l^*}+\frac{1-p_{k,z,l^*}}{N_k} )\sigmahat_{k,z,l^*}^2-(p_{k,z,r^*}+\frac{1-p_{k,z,r^*}}{N_k})\sigmahat_{k,z,r^*}^2\right).\label{eq:var-change} 
\end{align}

Note, we use true probabilities in \eqref{eq:var-change} for correct estimation of reduced variance as they can be computed using the whole data once the split point is known.
Since the strata are non-overlapping, $p_{k,z} = p_{k,z,l^*} + p_{k,z,r^*}$ and $N_{k,z} = N_{k,z,l^*} + N_{k,z,r^*}$. 
We define 
\begin{equation*}
    \delta_{k}(z) := \frac{\sigmahat^2_{k} - \sigmatilde_{k}^2(z)}{\sigmahat^2_{k}}, \label{eq:CART_red}
\end{equation*} as the proportion of variance reduced when node $z$ is split.
Now, the question is whether this reduction in variance is enough to split the node. To answer that, we define 
\begin{equation}
    G_k(z) :=  -\delta_k(z) \ln(\delta_k(z)),\label{eq:CART_modIG}
\end{equation}
which can be viewed as the information gained by splitting the node $z$ at $\theta_k$. Assuming that the split leads to a reduction in the estimated variance, $\delta_{k}(z) \in (0, 1)$, and hence the maximum theoretical value of $G_k(z) = 1/e$. This information gain value is computed for each leaf. Let $G_{k}^{\text{prev}}$ be the gain from the most recent accepted split. Then, the leaf selected for splitting is the one that maximizes $G_k(z)$ subject to providing a gain that is at least as good as the previous gain $G_{k}^{\text{prev}}$, i.e.,
\begin{align}
     \underset{\begin{subarray}{c} {z =1,2,\cdots,Z_k}
    \end{subarray}}{\max} &{G_k(z)},\nonumber\\
    \text{subject to: }&G_k(z) > G_{k}^{\text{prev}}.\label{eq:accept-split}
\end{align}

The selected leaf is then indexed appropriately, and its gain updates the value of $G_{k}^{\text{prev}}$. If~\eqref{eq:accept-split} has no feasible solutions, the splitting stops, and the tree and stratification structure is finalized. We also use another hyperparameter ($\tau$) common for binary trees that removes leaves with less than a certain number of data points sampled as shown in Algorithm~\ref{alg:synch_cart}. 
Generally, $G_k(z)$ is initially small, then it reaches the maximum value after the first few splits and starts reducing after that. The algorithm stops when $G_k(z)$ is close to the maximum value because finding a split that results in more gain is difficult after that. Note, another advantage of using the information gain strategy for splitting is that we can select which of the leaves provides the best split rather than splitting leaves in the order of their indexing.

\begin{algorithm}
\caption{Determining Strata (multi-D) using Binary Trees at Iteration $k$} 
\label{alg:synch_cart}
\begin{algorithmic}[1]
\item \textbf{Input:} Current iterate $\theta_k$, minimum leaf size threshold $\tau$, and loss values computed at  $(x_j,y_j)  \in \mcS_k$ where $|\mcS_k|=\lambda_k$.
\item \textbf{Initialization:} 
\State \hspace{1 em} Compute the first split by solving~\eqref{eq:CART_Obj} to get $\mcX_{k,1,l^{*}},\mcX_{k,1,r^{*}}$ and compute $G_k(1)$. 
\State \hspace{1 em} Set $G_{k}^{\text{prev}}= G_k(1)$, $\mcX_{k,1}=\mcX_{k,1,l^{*}}$ and $\mcX_{k,2}=\mcX_{k,1,r^{*}}$.
\State \hspace{1 em} Set $\mcI_k=\{\mcX_{k,1},\mcX_{k,2}\}$, $\mcS_{k,1}=\mcS_{k,1,l^{*}}$, and $\mcS_{k,2}=\mcS_{k,1,r^{*}}$. 
\State \hspace{1 em} Set $Z_k=2$ and update $\sigmahat_k^2$ following~\eqref{eq:var-est}.
\While{true}
    \For{$z\in\{1,2,\ldots,Z_k:\ |\mcS_{k,z}| > 2\tau\}$}
    \State Compute optimal split in $z$-th leaf by solving~\eqref{eq:CART_Obj} to get $\mcX_{k,z,l^{*}},\mcX_{k,z,r^{*}}$.
    \State If $\min\{|\mcS_{k,z,l^{*}}|,|\mcS_{k,z,r^{*}}|\}>\tau$, compute $G_k(z)$ via~\eqref{eq:CART_modIG}, else set $G_k(z)=-\infty$.
    
 \EndFor
 \If{there is an acceptable split, i.e., optimization~\eqref{eq:accept-split} has a solution}
 
    \State Set the leaf that solves~\eqref{eq:accept-split} as $z^{\text{split}}$ and remove $\mcX_{k,z^{\text{split}}}$ from $\mcI_k$.
    \State Set $G_{k}^{\text{prev}}= G_k(z^{\text{split}})$, $\mcX_{k,z^{\text{split}}}=\mcX_{k,z^{\text{split}},l^{*}}$ and $\mcX_{k,Z_k+1}=\mcX_{k,z^{\text{split}},r^{*}}$.
    \State Set $\mcI_k=\mcI_k\cup\{\mcX_{k,z^{\text{split}}},\mcX_{k,Z_k+1}\}$, $\mcS_{k,z^{\text{split}}}=\mcS_{k,z^{\text{split}},l^{*}}$,  $\mcS_{k,Z_k+1}=\mcS_{k,z^{\text{split}},r^{*}}$.
    \State Set $Z_k=Z_k+1$ and update $\sigmahat_k^2$ following~\eqref{eq:var-est}.
 \Else
    \State break
\EndIf
 
\EndWhile
\item \textbf{Output:} Stratification structure $\mcI_k$ with $Z_k$ many strata and samples $\{\mcS_{k,z}\}_{z = 1}^{Z_k}$. 
\end{algorithmic}
\end{algorithm}

\subsection{Stratification using Concomitant Variables}
Trees can partition the input space with multiple variables simultaneously. Yet, their drawback is the greedy heuristic search that can have intense computation at every iteration and sensitivity to smaller subsets of $\lambda_k$ pilot samples that estimate the leaf statistics. They are susceptible to producing less effective strata in the early iterations where the pilot run is small. They fall short of the attractive feature of asynchronous strata using large quantities of data (without running simulations). 

We propose a second method to construct strata that will, to the extent possible, use large quantities of data while still enjoying dynamic stratification. To motivate this method, we review the two properties of an ideal stratification variable. First, since the basis of stratification is conditioning the simulation output, input variables are helpful given that their distributional behavior can be inferred in each defined stratum without much burden. Second, a good stratification variable is one with a strong correlation (linear dependence) with the simulated output. In fact, it is possible to derive closed-form boundaries for input variables with known or partially known distributions that are linearly dependent on the stochastic objective function values. 
Variables that are auxiliary to the stochastic objective function value and are generated during a simulation run can hence be used in service of variance reduction; we term these variables, the concomitant variables. Concomitant variables' use for constructing strata boundaries is reminiscent of control variates and exploiting their linear dependence with the random output of interest~\citep{wilson1984variance}. \cite{jaisha2023} use the derived closed-form boundaries using optimal allocation for a queuing problem whose total cost one wishes to minimize. Simulated data considered for this purpose either have known distributions (service times) or unknown distributions (waiting times). In both cases the amount of data is limited because it is what the simulation runs will generate, but the waiting time appears to be a better concomitant variable given its more direct linear relationship with the total cost.

In this paper, we extend this viewpoint (by using proportional allocation instead of optimal allocation for stability) to the data-driven calibration with two new considerations: a) besides the simulation-generated data, we have a vast amount of real (not simulated) input data that can be used rapidly without running simulations to construct the strata; and b) to choose among real or simulated variables the most linearly dependent with the objective function, we include a number of their nonlinear transformations as potential candidates to serve as the concomitant variable. If the concomitant variable is chosen to be among the real input data, then it would provide the same boundaries given a number of strata for any visited $\theta$. But dynamic stratification will be due to the choice of the variable and the number of strata that can change from one iterate to another. We next describe different parts of the new approach, as laid out in Algorithm~\ref{alg:cv-boundaries}. 
\begin{algorithm}
\caption{Determining Strata (1-D) using Concomitant Variables at Iteration $k$}
\label{alg:cv-boundaries}
\begin{algorithmic}[1]
\item \textbf{Input:} Current iterate $\theta_k$, maximum number of strata $Z_{\max}$, loss  computed at $(x_j,y_j) \in \mcS_k$ where $|\mcS_k|=\lambda_k$, and thresholds $\varepsilon = 10^{-6}$ and $\varrho = 10^{-1}$.
\item \textbf{Initialization:} Collect candidate concomitant variables from linear/nonlinear transformation of real or simulated data $\{C_k^1,C_k^2,\ldots,C_k^r\}$. Set $Z=1$.
\For{$i=1,2,\ldots,r$} \label{step:choose-begin}
\State Fit a weighted regression model $F(\theta_k,(X,Y)) ~ \alpha^{i}_k + \beta^{i}_k C^{i}_k + E_k^{i}$.
\State Estimate $\widehat{\Var}(E_k^{i})$ and $\widehat{\text{Corr}}(C_k^{i},E_k^{i})$.
\EndFor
\State Find $C_k:=C_k^{i^*}$ where $i^*=\arg\!\min_{i=1,2,\ldots,r}\{\widehat{\Var}(E_k^{i}):\ |\widehat{\text{Corr}}(C_k^{i},E_k^{i})|<\varrho\}$.\label{step:choose-end}
\For{$Z=2,\ldots,Z_{\max}$}
\If{distribution of $C_k$ is known}
\State Look up $c_1 < c_2 < \cdots < c_{Z-1}$ and set $c_0=-\infty,c_{Z}=\infty$.
\Else 
\State Set $c_0=-\infty,c_{Z}=\infty$ and $c_1', c_2',\cdots, c_{Z-1}'$ as $Z-1$ quantiles of data.\label{step:begin}
\Repeat
\State Set $c_{z}=c_{z}' \quad \forall z = 1,2,\dots,Z-1$.
\State Estimate $\mu_z = \mbE[C_k|c_{z-1} \leq C_{k} < c_z] \quad \forall z = 1,2,\dots,Z-1$.\label{step:mu}
\State Set $c_{z}' = (\mu_z + \mu_{z+1})/2 \quad \forall z = 1,2,\dots,Z-1$.  
\Until{$\|(c_{z}')_{z=1}^{Z}-(c_{z})_{z=1}^{Z}\|\leq\varepsilon$}\label{step:end}
\EndIf
\State Set $\mcI_{k,Z} = \{\mcX_{k,z}\}_{z=1}^{Z}$, where $\mcX_{k,z}=\{X:\ C_k(X)\in[c_z,c_{z+1})\}$.
\EndFor
\State Determine $Z_k = \underset{\begin{subarray}{c} {Z \in [2,Z_{\text{max}}]}
    \end{subarray}}{\arg\!\min} {\widehat{\text{Var}}\left(\fhat_\text{post}(\theta_k,\lambda_k|\mcI_{k,Z})\right)}$ via bootstrapping.
    
    \item \textbf{output:} concomitant variable $C_k$ with $Z_k$ many strata and structure $\mcI_k = \mcI_{k,Z_k}$.
\end{algorithmic}
\end{algorithm}

\paragraph*{Boundaries on a concomitant variable:} Suppose $C=C(X)$ is the concomitant variable used for stratification---a linear/nonlinear transformation of a real variable in our dataset or a simulated variable generated alongside the simulated outputs of interest. Optimal stratification structure involves the boundaries $$c_{0} < c_{1} < \dots < c_{Z_k},$$ that minimize the variance of the stratified sampling estimator to obtain the stratification structure $\mcI_k$. $c_{0}$ and $c_{Z_k}$ are the two extreme values for $C$, typically considered to be $\pm\infty$. Leveraging post-stratification, boundaries that minimize the variance can be derived using the following theorem:
\begin{theorem}[\cite{dalenius1951problem}]
    Suppose the linear regression relation $F = \alpha + \beta C + E$ holds with $\mbE[E]=0, \Var(E)=\sigma_E^2$, and $\text{Cov}(C,E)=0$. Suppose also that we have a total of $n$ samples and want $Z$ strata on $C$. Then we can minimize the post-stratified estimator's variance to order $n^{-1}$ by choosing the strata boundaries 
    \begin{equation}
        c_z = \frac{\mbE[C\mid c_{z-1} \leq C < c_{z}]+\mbE[C\mid c_{z} \leq C < c_{z+1}]}{2}\ \forall z=1,2,\cdots,Z-1.\label{eq:partition_bounds_eqn}
    \end{equation}
\label{thm:opt-boundaries}\end{theorem}
While the closed-form equation~\eqref{eq:partition_bounds_eqn} is recursive and appears complex, under known probability distribution of $C$, $c_z$ quantities can be exactly computed and are accessible in look-up tables for several distributions \citep{sethi1963note}. The standard normal case for $C$ is relevant in simulation studies, providing good approximations for standardized variables that are aggregated statistics common in many discrete-event models. If the concomitant variable has an unknown distribution, we can solve for the optimal boundaries using a (relatively fast) convergent fixed-point iterative method \citep{burden19852}; see Step~\ref{step:begin}-Step~\ref{step:end} in Algorithm~\ref{alg:cv-boundaries}. During these steps, if $C$ is among the simulated data, then the conditional means are estimated with $\lambda_k \ll n$ pilot runs. Big data has less leverage in this case, similar to the binary tree approach. If $C$ is among the real data, its conditional means can be approximated with rapid population statistics to compute boundaries. In both cases, strata are exactly or approximately computed for a fixed number of strata. 

\paragraph*{Choosing the concomitant variable.} During optimization, there may be different variables at different iterations that provide the most linear relationship with the outputs, i.e., $C_k$ such that $$F(\theta_k,(X,Y)) = \alpha_k + \beta_k C_k + E_k,$$ where $E_k$ is the stochastic residual satisfying the assumptions in Theorem~\ref{thm:opt-boundaries}. One can use transformations of the original real or simulated variables to find the desired linear relationship (either through some descriptive data analysis or by using expert knowledge about the model). In this paper, we propose gathering a list of transformed variables as candidates for the concomitant variable and fitting a  weighted least squares linear regression model for each candidate; see Step~\ref{step:choose-begin}--Step~\ref{step:choose-end} in  Algorithm~\ref{alg:cv-boundaries}. The weighted least squares regression is preferred over ordinary least squares to account for the outliers, heterogeneous variance, and the erratic behavior of the simulations \citep{holland1977robust}. We choose a $C_k$ that yields the smallest \emph{ratio of variance} $\Varhat(E_k)/\Varhat(F_k)$; if the residual has a smaller variance than the response, then inference about the mean of the residual will be more precise that the same inference about the mean of the response~\citep{smith1991post}. Importantly, making an inference about the population by training a regression model using the samples has the risk of incorrectly estimating the regression coefficients $\alpha_k,\beta_k$. The variance of the residuals is independent of the stratification structure yet affected by this erroneous estimation. We emphasize that these operations are relatively fast given the use of $\lambda_k$ samples for fitting a whole bunch of regression lines. This is in contrast to the geometric number of operations in the binary tree to find the optimal stratification structure.

\paragraph*{Choosing the number of strata.} The last challenge is finding the number of strata. 
We decide the number of strata by determining $Z_k$ to find the lowest variance. In other words, if $\mcI_{k,Z}$ denotes the stratification structure with $Z$ strata, we find
\begin{equation}
   Z_k = \underset{\begin{subarray}{c} {Z \in [2,Z_{\text{max}}]}
    \end{subarray}}{\arg\!\min} {\widehat{\text{Var}}\left(\fhat_\text{post}(\theta_k,N_k=\lambda_k|\mcI_{k,Z})\right)},\label{eq:stratification_objective}
\end{equation}
where $Z_{\text{max}}$ is the user-defined upper limit on the number of strata; \cite{cochran1977sampling} proves that $Z_{\text{max}}=6$ for regression models with $R^2<0.95$. To evaluate the variance given $Z$ strata, we use $n_{\text{boot}}$ bootstraps of $\lambda_k$ simulated data to evaluate the variance for each $Z$ and identify one that consistently yields the smallest variance.

\section{EXPERIMENTAL RESULTS}\label{sec:results}

We compare the proposed stratification methods (BT and ConV) to the corresponding trust-region-based method without stratification (NS) and a number of widely used global optimization methods, including Bayesian Optimization~\citep{frazier2018bayesian}, Simulated Annealing~\citep{prudius2012averaging}, and Random Search~\citep{andradottir2006overview}. Our numerical analysis spans Monte Carlo examples with various variance structures (heterogeneity) and input dimensions, queuing simulations, and a data-driven calibration case study with real data from an offshore wind farm. In all of these problems, the objective (loss) function is the mean squared error (MSE) as in ERM cases~\eqref{eq:erm}, quantifying the discrepancy of the simulated and observed data. By minimizing this MSE, we aim to calibrate the simulation model $h(\theta,X)$ by finding the optimal parameter value $\theta$. Each solver is run 20 independent times (20 macroreplications), given a fixed computational budget, to obtain a distribution of the optimal calibrated parameter values, starting from the same initial point, $\theta_0$. In each macroreplication, common random numbers (CRN) are used across solvers to enable reproducibility and sharper comparison. For BO, a combination of RBF and white kernel is used with expected improvement as the acquisition function. The RBF kernel is effective for building surrogate models in BO because it can approximate any function given enough data and is infinitely differentiable, ensuring that the surrogate model is very smooth. The scale parameter of the RBF kernel is tuned by maximizing the log-marginal likelihood.  For NS, the initial sample size is set to 80 to scale adaptively based on the variance of the estimates. In BT, we use the minimum leaf size threshold $\tau=5$ while building the trees and for ConV, the maximum number of strata used is $Z_{\text{max}}=4$.

For the trust-region methods, we make comparisons from these large scale experiments by reporting intermediate recommended solutions at different budget points to track the optimization trajectory. Aligned with the SimOpt library platform~\citep{eckman2023diagnostic}, we post-process these solutions; the objective function value at these intermediate solutions is estimated using a validation set that is $30\%$ of the total data sampled independent from the modeling set that generates the optimization trajectory. The post-estimated objective function values for each macro-replication of each solver are then aggregated to obtain the mean and $95\%$ confidence intervals (CI) to obtain progress curves per expended budget.

We first present the results for some numerical examples in Section~\ref{sec:res_num_exmp}, followed by the results for queuing calibration in Section~\ref{sec:res_mm1}. In Section~\ref{sec:res_ws}, we present the wind case study and present an in-depth analysis of the two stratification approaches. Finally, in Section~\ref{sec:res_discussion}, we discuss the scenarios in which each proposed method may perform better than the others.

\subsection{Numerical Examples: Static Monte Carlo Simulations} \label{sec:res_num_exmp}

To compare the proposed methods with the existing global search methods, we consider the following numerical examples with different characteristics:
\begin{enumerate}
    \item Example 1: Quadratic model with heterogeneous noise.
    \begin{itemize}
        \item Physical system: $Y = (X_1 - 2)^2 + (X_2 - 2)^2 + \epsilon$, with 
        $\epsilon ~ \mcN(0,|X_1X_2 - 2|)$.
        \item Computer model: $h(\theta,X) = (X_1 - \theta)^2 + (X_2 - \theta)^2 $.
    \end{itemize}
    \item Example 2: Highly nonlinear model with homogeneous noise.
    \begin{itemize}
    \item Physical system: $Y = (X_1 - 2)^7 + (X_2 - 2)^2 + \epsilon$, with 
    $\epsilon ~ \mcN(0,1)$.
    \item Computer model: $h(\theta,X) = (X_1 - \theta)^7 + (X_2 - \theta)^2$.
    \end{itemize}
    \item Example 3: Highly nonlinear model with homogeneous noise and significant difference in scale of one variable against another.    \begin{itemize}
        \item Physical system: $Y = 1000(X_1 - 2)^5 + (X_2 - 2)^2 + \epsilon$,  
        with $\epsilon ~ \mcN(0,1)$.
        \item Computer model: $h(\theta,X) = 1000(X_1 - \theta)^5 + (X_2 - \theta)^2$.
    \end{itemize}
    \item Example 4: Homogeneous noise with interaction terms.
    \begin{itemize}
        \item Physical system: $Y = 2X_1X_2 + \epsilon$, 
        with $\epsilon ~ \mcN(0,1)$.
        \item Computer model: $h(\theta,X) = \theta X_1 X_2$.
    \end{itemize}
\end{enumerate}
In all the examples above, $X=(X_1,X_2)$ where $X_1 ~ U(0,4)$, $X_2 ~ U(0,4)$, and the optimal parameter value $\theta^*$ minimizes   $\|h(\theta,X) - Y\|_2^2$. A dataset of 1,000 data points in each macroreplication is generated and divided into modeling and validation sets with CRN. The optimal parameter for all the examples is 2, and the total budget for all solvers is 1,000 simulation runs. Utilizing the entire dataset to evaluate the objective function for a single $\theta$ would exhaust the entire budget. Therefore, to enable a comparison with BO, instead of using the entire data for evaluation in each round, we chose a random sample of 50 points for each objective function evaluation. We use the same $50$ samples in SA and RS as well. Additionally, in the BO implementation, given that the calibration parameter is one-dimensional, the initial surrogate model is built using a set of 10 randomly selected design points following the recommendation in \cite{loeppky2009choosing}, which leads to 10 total BO iterations. SA and RS each ran for 20 iterations. For BT, $X_1$ and $X_2$ are considered for stratification. The set of potential concomitant variables considered for ConV are  $\{X_1, X_2, X_1^2, X_2^2, X_1^3,X_2^3\}$. Since these concomitant variables are chosen from raw data, the approach is denoted as ConV-R.

\begin{figure}[H] 
\centering
\subfloat[][Example 1.\label{fig:JoS_num_exmp1}]{\includegraphics[width = 0.5\textwidth]{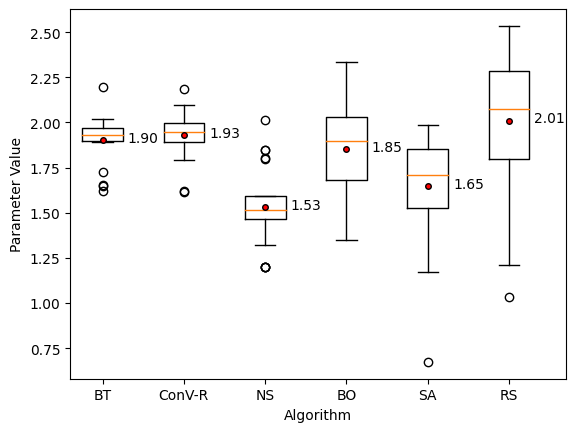}} 
\subfloat[][Example 2.\label{fig:JoS_num_exmp2}]{\includegraphics[width = 0.5\textwidth]{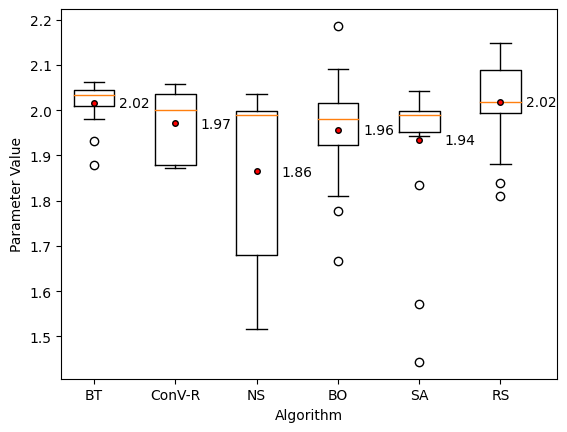}} \\
\subfloat[][Example 3.\label{fig:JoS_num_exmp3}] {\includegraphics[width = 0.5\textwidth]{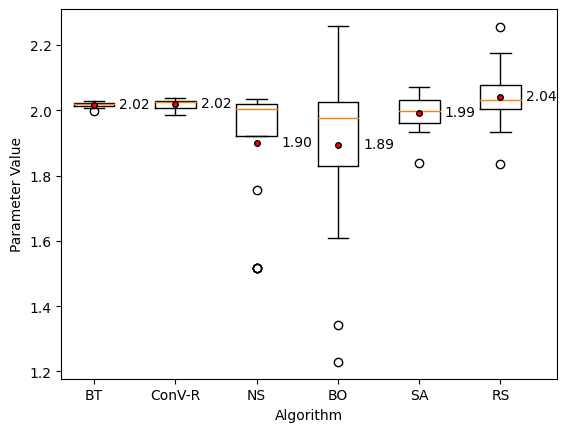}} 
\subfloat[][Example 4.\label{fig:JoS_num_exmp4}] {\includegraphics[width = 0.5\textwidth]{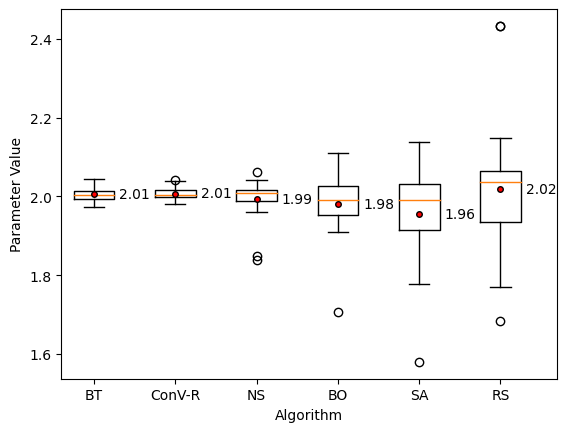}} 
\caption{Distributions (box plots) of the calibrated parameter values from 20 independent runs of the algorithms in numerical examples. The red dot and the numerical value along each box display the mean value. BT and ConV perform similarly and better than other solvers in mean value and more concentrated distribution. BT exhibits less variability in Examples 1 and 2.}
\label{fig:JoS_num_exmps}
\end{figure}

Figure~\ref{fig:JoS_num_exmps} illustrates how the final solutions vary across the 20 independent runs for each algorithm. Given the limited budget, BT and ConV-R consistently identify the optimal parameter value $\theta^*=2$ with minimal variability. In contrast, NS (the same adaptive solver but without stratification) and the global search methods exhibit high variability and sometimes fail to return near-optimal solutions. The high variability implies high risk of these solvers, in the sense that, even if the mean calibrated value is close to the optimal parameter value, a single run of these solvers is more likely to produce a suboptimal solution compared to the proposed approaches. While global methods are expected to converge to the optimal solution with a sufficient computational budget, this experiment shows that the proposed methods can achieve near-optimality faster and more consistently.

Additionally, the local search method without stratification is also prone to high variability or slow convergence compared to the dynamically stratified BT and ConV. BT shows slightly less variability between the two proposed methods in Example 1 and better mean and variance in Example 2. This evidence suggests that BT may be more robust in the presence of heterogeneous noise and more extreme nonlinearity. However, for the latter case, ConV may perform better with the additional pre-processing to nonlinearly transform $X_1$ and $X_2$.

\subsection{Calibrating a Queuing Model: Discrete-Event Simulation Example} \label{sec:res_mm1}

In this example, we use calibration to determine the optimal interarrival rate in a simple M/M/1 queue given synthetic data comprising mean service time, mean waiting time, and mean sojourn time. We minimize the discrepancy of the simulated mean waiting time $h(\theta,X)$ (using interarrival rate $\theta$) and the observed mean waiting time $Y$.

The distinction of this example with the static simulations in the previous section is that for ConV method, here we can aggregate a sequence of random variables generated over time and apply Central Limit Theorem (CLT) via standardized mean service time and standardized mean sojourn time as potential concomitant variables \citep{wilson1984variance}. The advantage of this transformation is that the ConV method can leverage a lookup table for the optimal strata boundaries of the standardized variables~\citep{jain2023post, sethi1963note, wilson1984variance}  and bypass any additional computation  (Step~\ref{step:begin}--Step~\ref{step:end} in Algorithm~\ref{alg:cv-boundaries}). As a result, more precision and less clock-time computation in ConV method compared to the BT method may be achieved.

In the experiment, we use a synthetic dataset of 10,000 observations. Each macroreplication starts at the same starting point $\theta_0 = 1.5$ and has a total budget of 10,000 simulation runs; similar to the previous section,  $30\%$ of the dataset in each macro-replication is randomly held for post-processing. The length of the discrete-event queuing simulation is 200 with a warmup period of 50 to obtain a steady state. The service rate is 2, and the \emph{optimal} interarrival rate (used to generate the waiting and sojourn times in the dataset) is 1. The standardized mean service time and standardized mean sojourn time are utilized as stratification variables for ConV-R. For BT, the mean service time and mean sojourn time are used for stratification. While implementing BO, SA, and RS for comparison, a random sample of 200 points were used in each objective function evaluation.

\begin{figure}[H] 
\centering
\subfloat[][95$\%$ CI of progress curves generated from the proposed approaches for the queuing calibration reveals that ConV-R performs significantly better than BT with lower optimal loss and much less variability.\label{fig:JoS_mm1_cc}]{\includegraphics[width = 0.49\textwidth]{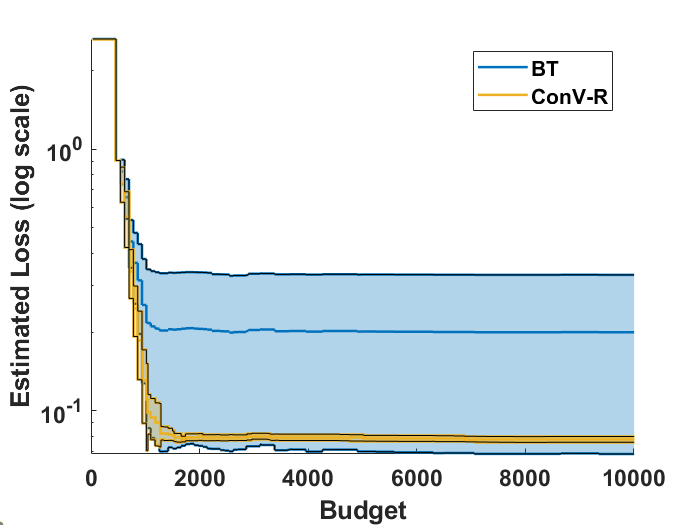}} \hfill
\subfloat[][Distribution of the calibrated parameter values from 20 independent runs of the algorithm for the queuing calibration. The red dot and the numerical value along the boxplot display the mean value. BT and ConV-R are more accurate with smaller variances.\label{fig:JoS_mm1_bp}]{\includegraphics[width = 0.49\textwidth]{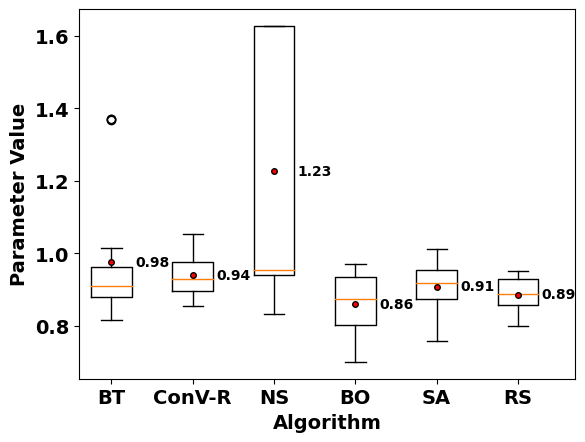}} \\
\caption{Comparison of the performance of solvers for the queuing model calibration.}
\end{figure}

Figure~\ref{fig:JoS_mm1_cc}
illustrates the evolution of the objective function across 20 macroreplications as the percentage of expended simulation budget increases. 
The performance of ConV-R is better than BT as it achieves a lower MSE with minimal variability across macroreplications. This superior performance is anticipated, given that the asymptotic normality of the stratification variables simplifies the implementation of ConV-R and allows for the use of exact strata boundaries derived from theoretical principles without relying on assumptions or error-prone estimations. Importantly, we observe in Figure~\ref{fig:JoS_mm1_bp}  that the mean of optimal solutions recommended by BT is closer to the optimal parameter $1$; however, the outlier optimal solutions (in the boxplot) and larger variability correspond to significantly worse objective function values, which is depicted by Figure~\ref{fig:JoS_mm1_cc}. The very small variability of the optimal loss values returned by ConV's distribution of optimal solutions suggests that despite their variability, they all yield similar small loss values. Another important observation from Figure~\ref{fig:JoS_mm1_bp} is the poor performance of other solvers, above all the NS (same adaptive solver but without stratification) in calibrating the queuing model.

\subsection{Wind Case Study: Wake Model Calibration} \label{sec:res_ws}

\begin{figure}[H]
 \centering
 \includegraphics[width = 0.6\textwidth]{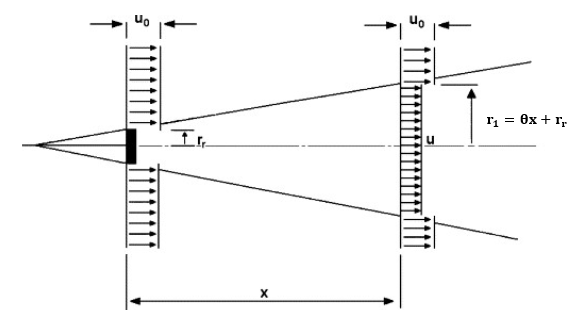}
 \caption{Linear propagation of wake as modeled by the Jensen wake model, where $r_r$ is the rotor radius and $u_0$ is the free-stream wind speed (excerpted from \cite{jensen1983note}).}
 \label{fig:Jensen_wake}
\end{figure}

Recall the example we started with in Section~\ref{sec:Intro}. The \emph{wake} effect causes the wind speed reaching the downwind turbines to be less than the wind speed at the upwind turbines, affecting the power generated by these turbines~\citep{you2017wind}. Jensen wake model \citep{jensen1983note} is a simple but widely used wake model extendable to a multi-turbine setting \citep{katic1986simple} that assumes that wake propagates linearly in the downwind direction, as shown in Figure~\ref{fig:Jensen_wake}. The value of the wake decay coefficient ($\theta$ in Figure~\ref{fig:Jensen_wake}) impacts the performance of the Jensen wake model. Though a value of $\theta = 0.04$ is widely assumed 
for offshore wind farms \citep{barthelmie2010quantifying,katic1986simple}, some recent studies have shown that this value does not necessarily depict the wind speed reduction observed in actual wind farms \citep{goccmen2016wind,you2017direction}. Thus, it is essential to determine the value of this wake decay coefficient for each wind farm separately. The wake model simulates the wind speed at each turbine in the wind farm. The power curve for the turbines is generated via B-splines by using the data at one of the upwind turbines \citep{lee2013bayesian,you2017direction}. This power curve is used to estimate the power generated at each turbine.

In our case study, data is collected from an offshore wind farm with 30+ turbines. The data includes information about wind conditions, such as the 10-minute average wind speed (WS) and direction, turbulence intensity (TI), etc. Along with this, it also consists of a 10-minute average power generated by each turbine. In the analysis, the power generated by each turbine is normalized by dividing it by the maximum possible power that can be generated, referred to as nominal power \citep{milan2010power,byon2011simulation}. For a given combination of input wind condition $X$ (WS, TI, etc.) and the wake decay coefficient $\theta$, Jensen's wake model estimates the power generated by turbines $h(\theta; X)$. This simulated power is then compared to the observed power at turbines $Y$ to get $F(\theta,(X,Y))$, the objective function value, the loss function measuring the discrepancy between observed power and model predicted ones.

\subsubsection{Implementation}

A modeling set comprising $70\%$ data used for optimization is sampled independently for each macro-replication. Each macro-replication starts at the same initial point $\theta_0 = 0.1$, the initial TR radius $\Delta_0= 0.08$ and the minimum sample size $\lambda_0 = 80$, and runs for a total of 10,000 simulations (budget). In our first proposed approach, the input variables WS and TI are used as the stratification variables for dividing the data via binary trees (\textit{BT}).  

In our second proposed approach, two cases are considered for stratification with concomitant variables: using real data (\textit{ConV-R}) or the simulated data (\textit{ConV-S}). In the first method, we consider five alternatives to stratify the data using input WS and TI along with nonlinear transformations $\text{WS}^2$, $\text{TI}^2$, and $\text{WS}^3$. With unknown joint probability distribution of TI and WS, the strata boundaries are determined by solving the iterative method using the population $\mcX$. Based on the stratification boundaries, the real data is divided into non-overlapping sets $\mcX_{k,z},\ z=1,2,\ldots,Z_k$, and the probabilities are $p_{k,z} = |\mcX_{k,z}|/|\mcX|$. For a given $\theta_k$, the Jensen model simulates the wind speeds reaching each turbine, providing the mean estimated wind speed at the turbines  $\widehat{\text{WS}}_k$. The model also provides the simulated power at each turbine using this simulated wind speed and the power curve. Thus, another possibility of a concomitant variable is the mean estimated power at the turbines $\hat{h}(\theta_k,X)$. For stratification using simulated data (\textit{ConV-S}), we consider these two variables along with their nonlinear transformations $\widehat{\text{WS}}_k^2, {\hat{{h}}}^2(\theta_k,X),$ and $\widehat{\text{WS}}_k^3$. When using these simulated variables for stratification, the strata boundaries are determined by using the iterative method with $\lambda_k$ points, and the probabilities are estimated as $p_{k,z} \approx \lambda_{k,z}/\lambda_k$. For both $\textit{ConV-R}$ and $\textit{ConV-S}$ at each iteration, the variable with the lowest residual variance is chosen as the concomitant variable. Thus, we do not choose a concomitant variable a priori; the algorithm identifies it adaptively.

\subsubsection{Results}

\begin{figure}[H] 
\centering
\subfloat[][Progress of the mean objective function value during optimization over 20 macro-replications for $\theta_0 = 0.1, \Delta_0 = 0.08$, and $\lambda_0 = 80$. The x-axis represents the number of simulations.\label{fig:JoS_mean_CI_stratification_vs_no_stratification}]{\includegraphics[width = 0.49\textwidth]{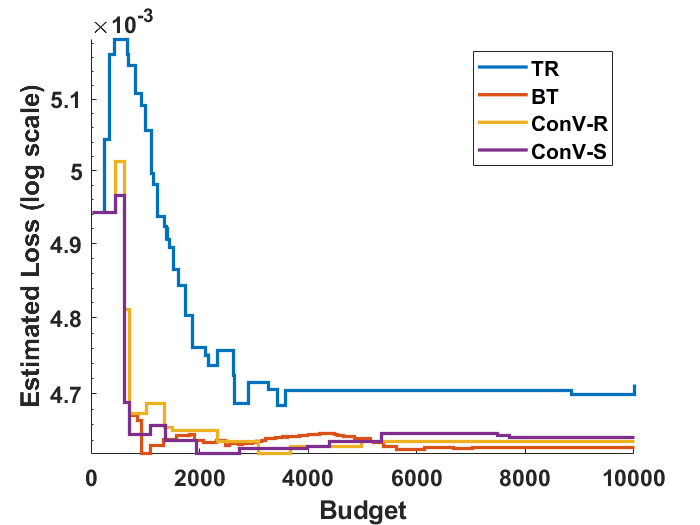}} \hfill
\subfloat[][Distribution of the calibrated parameter values from 20 independent runs of the algorithm for the wind case study. The red dot and the numerical values along the boxplot displays the mean value.\label{fig:JoS_ws_bp}]{\includegraphics[width = 0.49\textwidth]{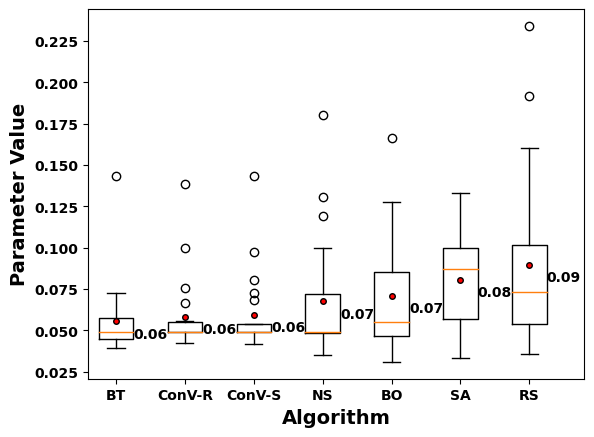}} \\
\caption{Comparison of the performance of solvers for the wake model calibration. BT and ConV perform better than the global solvers.}
\end{figure}

\begin{figure}[h] 
\centering
\subfloat[][Binary trees.\label{fig:JoS_CI_area_a}]{\includegraphics[width = 0.35\textwidth]{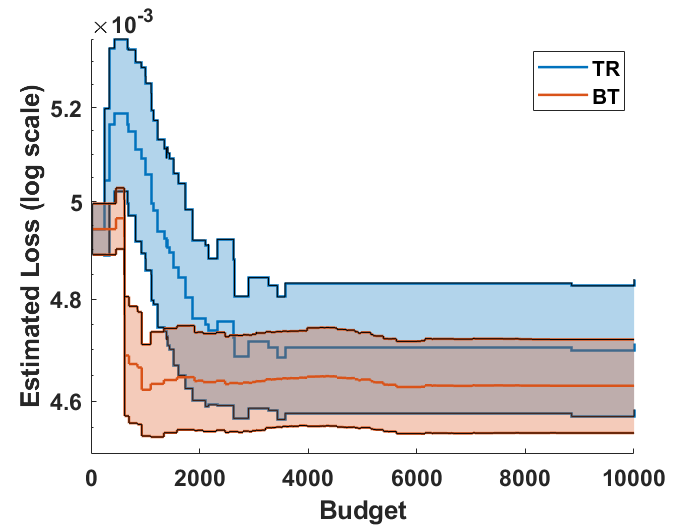}} 
\subfloat[][Concomitant variables: real data.\label{fig:JoS_CI_area_b}]{\includegraphics[width = 0.35\textwidth]{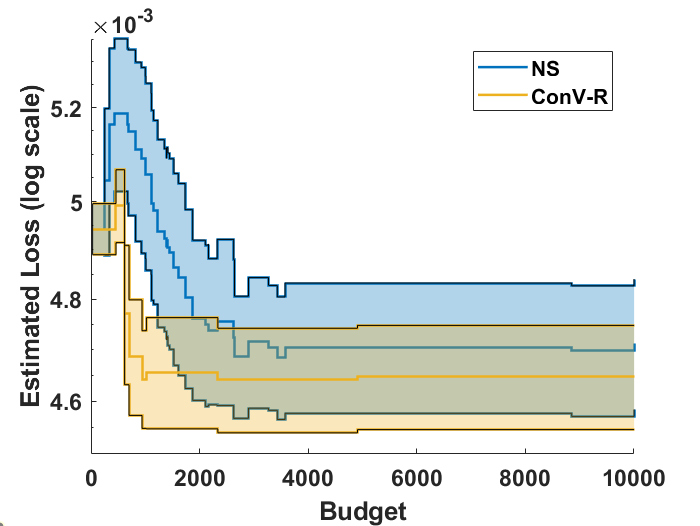}} 
\subfloat[][\centering Concomitant variables: simulated data.\label{fig:JoS_CI_area_c}]{\includegraphics[width = 0.35\textwidth]{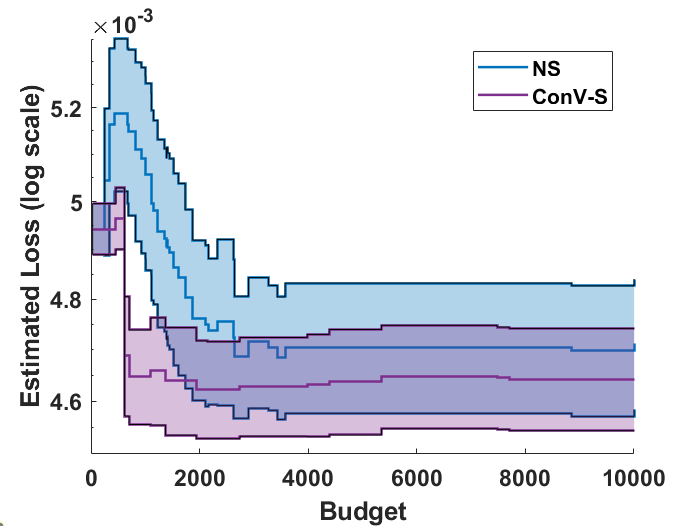}}
\caption{Variability and risk ($95\%$ CI progress curves) of proposed approaches (\textit{BT, ConV-R,} and \textit{ConV-S}) is reduced compared to no-stratification (\textit{NS}), computed over 20 macro-replications.}
\label{fig:JoS_CI_area}
\end{figure}

Figure~\ref{fig:JoS_mean_CI_stratification_vs_no_stratification} compares how the expected progress varies during optimization for the no stratification case (\textit{NS}) and the solvers with dynamic stratification all under ASTRO-DF optimization algorithm. The main advantage of using the proposed stratified adaptive methods is a significant improvement in performance initially to reach better solutions. All of the three proposed approaches provide comparable results. \textit{BT} and \textit{ConV-R} exhibit more similar performance, which is interesting given that \textit{ConV-R} uses only one variable at a time for stratification. Another observation is that they both reach better solutions compared to \textit{ConV-S}, which is expected as \textit{ConV-S} builds the stratification structure with a small sample of noisy simulated data. The second advantage is depicted by Figures~\ref{fig:JoS_CI_area_a},~\ref{fig:JoS_CI_area_b}, and~\ref{fig:JoS_CI_area_c} where the 95$\%$ CI widths of \textit{BT}, \textit{ConV-R}, and \textit{ConV-S} are smaller, indicating reduced variability or uncertainty (risk) in the performance of the optimization algorithm. 

Figure~\ref{fig:JoS_ws_bp} compares the performance of  BT and ConV against the global search algorithms (BO, SA, and RS) and the not-stratified version of the adaptive local solver (NS). The stratification methods show much lower variability. Although the mean calibrated parameter values for BT and ConV are similar, BT exhibits more variability (larger interquartile range).

\begin{table}[h]
    \centering
    \caption{Mean frequency with which a particular variable is picked for stratification across 20 macro-replications. Note that these values are for $\theta_0 = 0.1, \Delta_0 = 0.08$, and $\lambda_0 = 80$. The distribution can change with the changes in these initial settings (the value in the parenthesis is the standard error).}
    \small
    \begin{tabular}{ccccccc}
    \hline
         \multirow{2}{*}{\textit{ConV-R}} & Variables &  WS & TI & $\text{WS}^2$ & $\text{TI}^2$ &$\text{WS}^3$\\ \cmidrule{2-7}
         & Frequency & 0.05(0.05) & 11.20(2.36) & 0.35(0.30) & 24.50(3.48) & 3.00(1.55) \\ \hline \hline
        \multirow{2}{*}{\textit{ConV-S}} & Variables &  $\widehat{\text{WS}}_k^2$ & $\hat{h}(\theta_k,X)$ & $\widehat{\text{WS}}_k^2$ &  ${\hat{{h}}}^2(\theta_k,X)$ &$\widehat{\text{WS}}_k^3$\\ \cmidrule{2-7}
         & Frequency & 2.00(1.05) & 0.30(0.22)  & 0.20(0.14) & 32.50(2.51) & 1.10(0.45) \\
         \hline
    \end{tabular} 
    \label{tab:variable_freq}
\end{table}

Recall that \textit{ConV-R} and \textit{ConV-S} dynamically identify the best concomitant variable throughout iterations. Table~\ref{tab:variable_freq} summarizes the mean frequency with which a concomitant variable is chosen for the baseline case $\theta_0 = 0.1, \Delta_0 = 0.08$, and $\lambda_0 = 80$. For \textit{ConV-R}, TI and its squared transformation are often picked for stratification. In the literature, the wake decay coefficient has been shown to correlate well with TI \citep{barthelmie2015role,duc2019local,pena2016application}. Thus, consistently choosing TI by the algorithm indicates that it is aptly choosing the best stratification variable. In \textit{ConV-S}, the mean of the squared simulated power at each turbine is chosen almost every time. The loss function is mathematically more correlated with ${\hat{{h}}}^2(\theta_k,X)$ than any other variable. Hence, choosing it consistently again indicates that the proposed method can determine the best stratification variable.

\begin{figure}[H] 
\centering
\subfloat[0$\%$ budget, $\theta_k = 0.100$.\label{fig:JoS_CART_evolving_strata_a}]{\includegraphics[width = 0.35\textwidth]{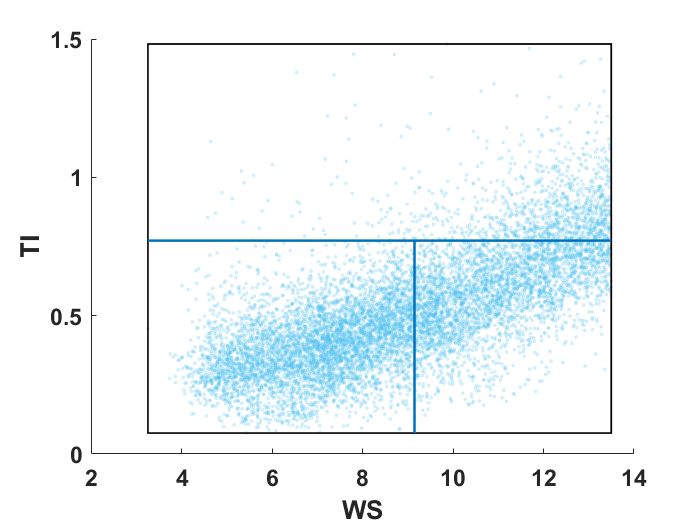}}
\subfloat[50$\%$ budget, $\theta_k \approx 0.049$.\label{fig:JoS_CART_evolving_strata_b}]{\includegraphics[width = 0.35\textwidth]{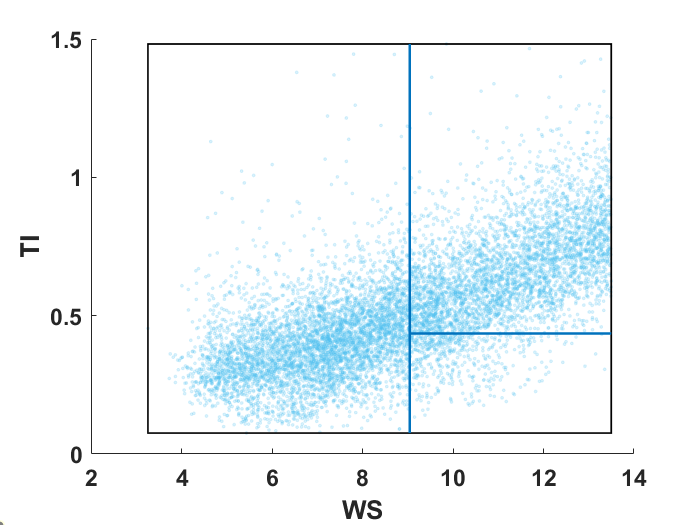}} 
\subfloat[100$\%$ budget, $\theta_k \approx 0.044$.\label{fig:JoS_CART_evolving_strata_c}]{\includegraphics[width = 0.35\textwidth]{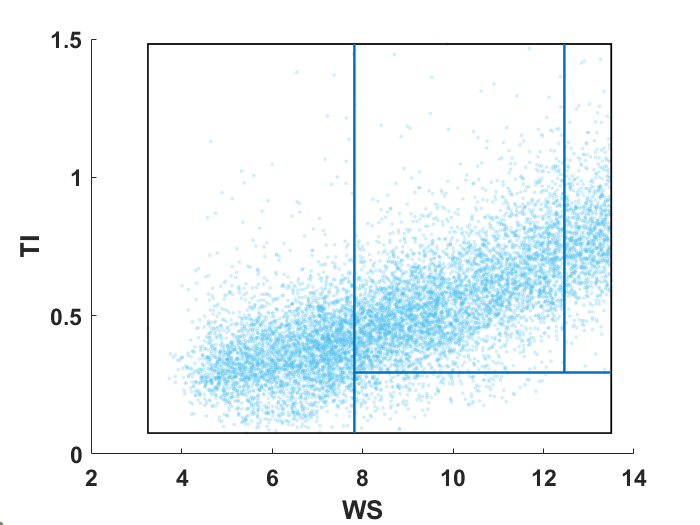}}
\caption{Evolution of the stratification structure, within \textit{BT}, during optimization for a single macro-replication after approximately 0$\%$, $50\%$, and $100\%$ budget is utilized. The points denote the actual data.}
\label{fig:JoS_CART_evolving_strata}
\end{figure}

\begin{figure}[H] 
\centering
\subfloat[][0$\%$ budget, $\theta_k = 0.100$.\label{fig:JoS_CVIn_evolving_strata_a}]{\includegraphics[width = 0.35\textwidth]{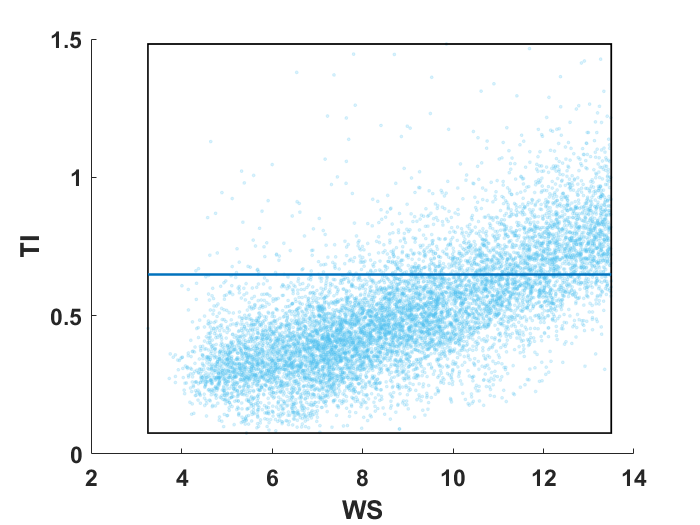}}
\subfloat[][50$\%$ budget, $\theta_k \approx 0.003$.\label{fig:JoS_CVIn_evolving_strata_b}]{\includegraphics[width = 0.35\textwidth]{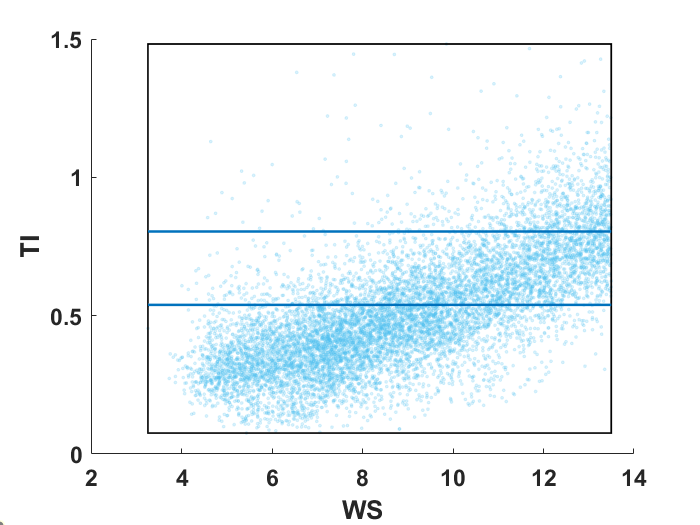}} 
\subfloat[][100$\%$ budget, $\theta_k \approx 0.049$.\label{fig:JoS_CVIn_evolving_strata_c}]{\includegraphics[width = 0.35\textwidth]{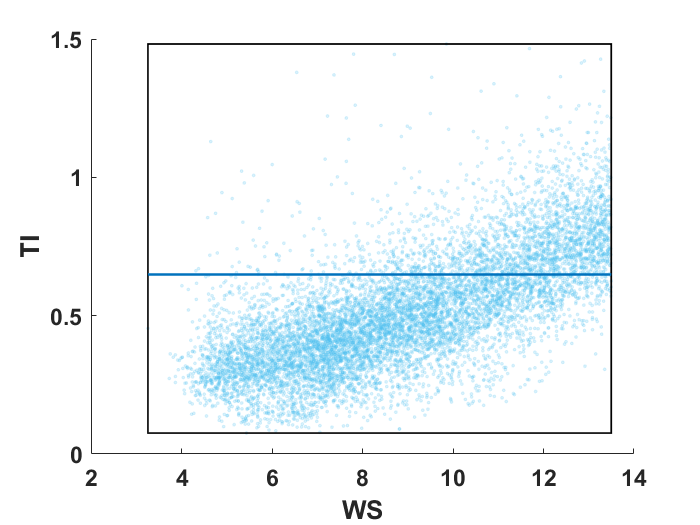}}
\caption{Evolution of the stratification structure, within \textit{ConV-R}, during optimization for a single macro-replication after approximately 0$\%$, $50\%$, and $100\%$ budget is utilized. The points denote the actual data.}
\label{fig:JoS_CVIn_evolving_strata}
\end{figure}

Figures~\ref{fig:JoS_CART_evolving_strata} and~\ref{fig:JoS_CVIn_evolving_strata} show how the stratification structure changes during optimization when using \textit{BT} and \textit{ConV-R} respectively. Unlike stratification with concomitant variables, binary trees can divide the data based on multiple variables (TI and WS), as shown in Figures~\ref{fig:JoS_CART_evolving_strata_a},~\ref{fig:JoS_CART_evolving_strata_b}, and~\ref{fig:JoS_CART_evolving_strata_c}. While computationally more intensive, this method is more flexible in choosing the stratification variable and deciding the number of strata. Recall that in \textit{BT}, real data corresponding to $\lambda_k$ pilot simulations is used for stratification, and in \textit{ConV-R}, the entire data is used for stratification. If the number of strata and the stratification variable are the same, \textit{ConV-R} will generate the same structure irrespective of $\theta_k$ as seen in Figure~\ref{fig:JoS_CVIn_evolving_strata} where the strata design for $\theta_k = 0.100$ and $\theta_k=0.049$ is the same. However, the number of strata and the stratification variable depends on $\theta_k$, which makes the stratification dynamic in \textit{ConV-R}. Additionally, the choices for the stratification structure throughout the optimization are finite (for each possible concomitant variable and each possible number of strata), which can reduce the run-to-run variability of the algorithm compared to other cases where there may be virtually infinite choices for the stratification structure.

\begin{figure}[H] 
\centering
\subfloat[][Influence of $\theta_0$.\label{fig:JoS_SA_theta}]{\includegraphics[width = 0.35\textwidth]{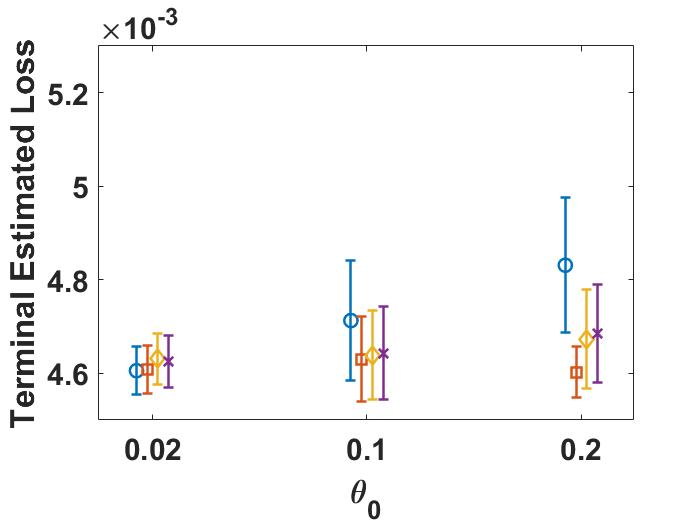}}
\subfloat[][Influence of $\Delta_0$.\label{fig:JoS_SA_Delta}]{\includegraphics[width = 0.35\textwidth]{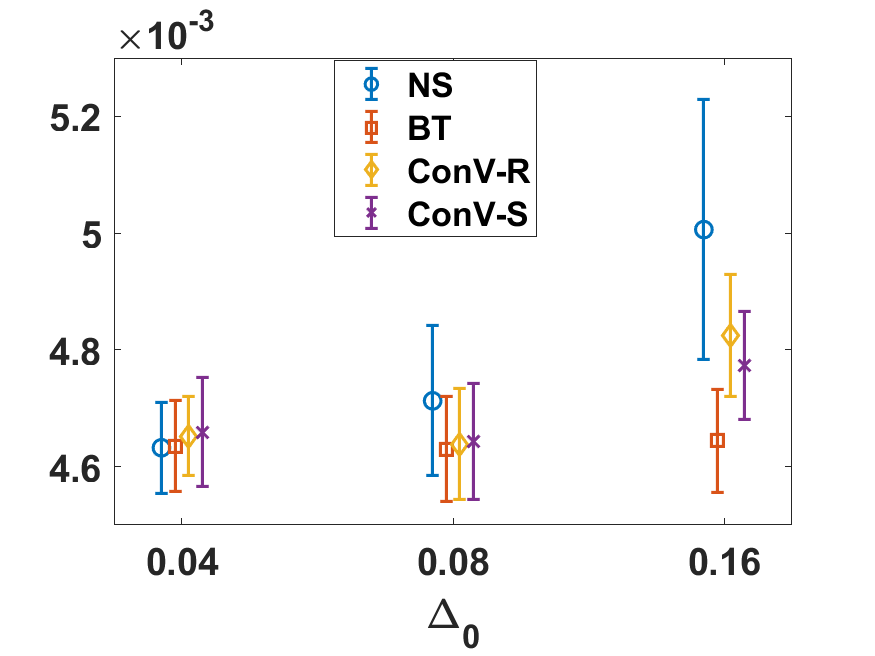}}
\subfloat[][Influence of $\lambda_0$.\label{fig:JoS_SA_lambda}]{\includegraphics[width = 0.35\textwidth]{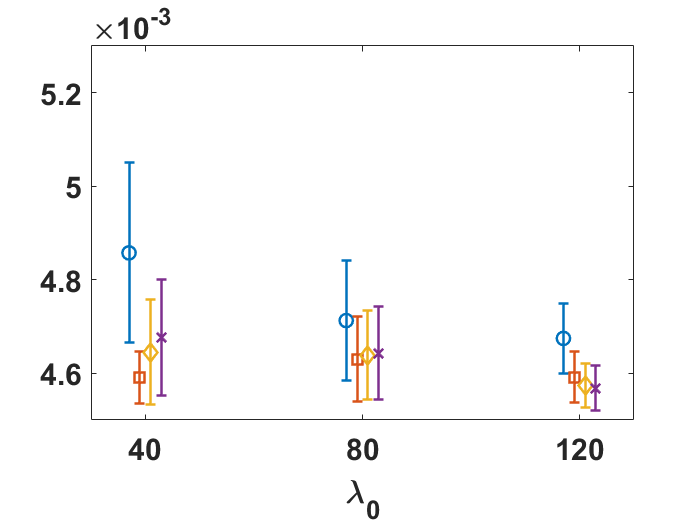}}
\caption{Effect of different hyperparameters on the performance of the solvers. Implementing stratification reduces the algorithm's dependence on the choice of the hyperparameters (Baseline setting: $\theta_0 = 0.1, \Delta_0 = 0.08,$ and $\lambda_0 = 80$).} 
\label{fig:JoS_SA}
\end{figure}

Next, we compare the robustness of the proposed methods with a Box-Wilson Central Composite Design (CCD). A CCD provides enough information to estimate the main effects and interactions with significantly fewer designs than a full-factorial design~\citep{hill1966review}. We test the proposed methods' sensitivity by varying the algorithm's three most critical hyperparameters, $\theta_0, \Delta_0,$ and $\lambda_0$. 
Considering $\theta_0 = 0.1, \Delta_0 = 0.08$, and $\lambda_0 = 80$ as the baseline case, the robustness is analyzed by fixing two parameters and perturbing the third between two relatively extreme values. We consider the following parameter values for the analysis: $\theta_0 = \{0.02, 0.2\}$, $\Delta_0 = \{0.04, 0.16\}$, and $\lambda_0 = \{40, 80\}$. Detailed hyperparameter tuning is beyond the scope of this paper; the ranges selected are reasonable for each parameter in the context of this problem and the objective of this sensitivity analysis study is to examine whether the performance of the proposed algorithms would significantly vary for different starting conditions. Figure~\ref{fig:JoS_SA} depicts the error bars of each solver's terminal objective function values obtained from 20 macro-replications, considering various designs. In summary, efficient dynamic stratification diminishes the reliance of TR algorithms on hyperparameters, enhancing their robustness.  

Figure~\ref{fig:JoS_SA_theta} investigates the influence of initial solution $\theta_0$ on the performance of the solvers. With a favorable starting point, $\theta_0 = 0.02$ (where we speculate that the objective function is at a steep region), the performance across all cases is about the same. This observation aligns with expectations, as the proximity of the starting point to the true optimum allows the algorithms to reach the optimal solution with minimal exploration. Conversely, for $\theta_0 = 0.2$, where the starting point is considerably far from the true optimum and at a more flat region, \textit{NS} exhibits significantly worse performance than using dynamic strata, highlighting that changing strata effectively can lead to robust exploration and, thus, better performance.

Figure~\ref{fig:JoS_SA_Delta} illustrates the effect of the initial TR radius, $\Delta_0$. A larger $\Delta_0$ facilitates early exploration and demands that the solvers execute efficient exploitation. Failure to accomplish this may lead to the algorithm becoming trapped in a suboptimal region. This is particularly evident in the case of \textit{NS}, where its performance deteriorates with increasing $\Delta_0$. In contrast, dynamic strata enhance early exploitation to a certain degree, enabling the solvers to attain improved solutions.  Stratification with concomitant variables shows some sensitivity to the initial TR radius, degrading their performance for larger $\Delta_0$ values. The enhanced flexibility provided by \textit{BT}, allowing the selection of multiple stratification variables simultaneously, may contribute to improved early exploitation, potentially explaining its performance for $\Delta_0 = 0.16$.

Figure~\ref{fig:JoS_SA_lambda} demonstrates how the initial sample size, $\lambda_0$, influences the solver's performance. A small $\lambda_0$ allows the algorithm more budget for exploration but can lead to imprecise estimates and, consequently, poor exploitation. As $\lambda_0$ increases, the performance of all solvers generally improves, but for a limited budget setting, a large $\lambda_0$ may not be preferable. Stratified sampling becomes crucial in this context as it provides more accurate estimates for smaller sample sizes. This capability allows solvers employing dynamic stratified sampling to outperform others, even when $\lambda_0$ is small. 
 
\subsection{Discussion} \label{sec:res_discussion}

While adaptive sampling in NS efficiently allocates the simulation budget for each 
$\theta$ based on its unconditional output variance and proximity to optimality, stratified sampling further enhances efficiency by reducing output variance with conditioning and allowing the sample size stopping conditions to be met earlier. The effectiveness of stratified sampling depends on the stratification structure, which optimally partitions the input space to minimize output variance.
Since output variance structure can significantly vary from one system (calibration parameter) to another---heteroscedasticity (with respect to $\theta$), the proposed dynamically stratified adaptive sampling procedure aims to learn about the local variance structure of the output to maximize efficiency. Any optimal budget allocation during optimization can enhance exploitation by improving estimation error with fewer samples (simulation runs) at each evaluation. Thrifty exploitation ensures ample budget is saved for exploration and allows the algorithm to run for more iterations. Therefore, a dynamically stratified adaptive sampling procedure increases the solver's ability to more thoroughly explore the decision space. 

In our experiments, the benefit of stratification is evident across all examples, consistently outperforming the non-stratified approach and the global solvers, as the recommended parameter values using dynamically stratified adaptive sampling are almost always closer to the true optimal and more consistent (less risky) across independent solver runs. Such advantages are notable, especially because the stratification procedure is not very time-consuming as the additional time needed for stratification is negligible compared to each simulation run. For example, in our queuing experiments, the average clock time to solve with NS (no stratification) was 142.0 seconds, while that of the BT and ConV solvers was 146.1 and 147.8 seconds, repectively. The rewards in finite-time solution performance easily justifies the added  $\sim 4\%$ increase in clock time. The computational overhead for stratification becomes more negligible when the simulation runtime is longer, such as wind power simulation.

Which dynamic stratification method should one choose? The answer to this question is contingent upon the structural characteristics of the problem. We highlight the following observations that can aid in incorporating these dynamically stratified adaptive sampling procedures within a solver: 
\begin{itemize}
    \item[(i)] While in most cases the two approaches perform similarly, Example 2 in Section~\ref{sec:res_num_exmp} suggests that BT may perform better than ConV in extreme nonlinearity
    or interactions (dependencies) in input variables. 
    \item[(ii)] For ConV to work well in these situations, additional pre-processing to find a good nonlinear transformation of the input variables may be necessary. This is because ConV relies on a linear mapping between what it will use as the concomitant variable and the objective (loss) function value $F$. At the minimum, the squared transformations of the input variables should be considered in calibration problems with MSE-like loss functions. How to find more linearly related concomitant variables inexpensively and whether that effort may be worthwhile is an open research question. 
    \item[(iii)] ConV has another restriction in only choosing one variable to stratify each time. At the time of writing this paper, we are unsure of whether that restriction necessarily translates to a weakness since in more extensive experimentation that we did not include in this paper, ConV performed competitively with BT for higher-dimensional and more inter-dependent input spaces or cases where multiple stratification variables held significance. An explanation for this observation may be that by the parsimony principles~\citep{goloboff2003parsimony}, finding the single input variable that is the major contributor to heterogeneity of output variance in the input space for a fixed $\theta$ may be sufficient and less prone to statistical errors and biases when forming the strata. This point is visible in the wind power calibration case study in Figure~\ref{fig:JoS_ws_bp}, where the BT boxplot of optimal solutions is wider compared to ConV and we see in Figure~\ref{fig:JoS_CART_evolving_strata} that BT sometimes stratifies with more than one input variable 
    \item[(iv)] Our sensitivity analysis with real data suggests that BT can have a slightly more robust performance compared to ConV with respect to the solver's starting conditions (initial solution, minimum sample size, initial step size); yet a more extensive experimental design to make a general judgement on sensitivity is left for future research.
    \item[(v)] ConV becomes efficient when the distribution of the stratification variable is either known or can be approximated, as seen in Section~\ref{sec:res_mm1} where the distribution of standardized mean service time and the standardized mean sojourn time can be approximated. For most of the time-dependent simulations, it is easy to use CLT to approximate the distribution of variables, and for these cases, stratification with concomitant variables can be very effective. 
    \item[(vi)] In using ConV, if the concomitant variables are chosen from the real (not simulated) input data, establishing strata requires minimal computation. More importantly, using all of the real data does not affect the simulation budget and significantly reduces the inherent noise when statistics from each strata is used to estimate the objective function or the sample size. 
\end{itemize}

\section{CONCLUSION}\label{sec:conclusion}
In data-driven calibration, the presence of abundant data with many covariates can help match the model outputs and observed outputs by tuning the model parameters. To reduce computation, using subsamples of data makes the problem stochastic and apt for simulation optimization, in which a vast amount of joint information can aid using stratified sampling to reduce estimation error at each visited calibration parameter. However, stratified sampling within simulation optimization is challenging. We propose using post-stratification to lower the instability of sampling distributions throughout the optimization. This stability enables a more tailored design of strata that, if done at a low cost, has the potential to save exploitation sampling efforts for more exploration in the search.

We further propose two ways for dynamic stratification. The first approach determines strata boundaries by hierarchically dividing the data using binary trees that are grown only until enough information can be gained. This approach may be computationally expensive but is flexible as it can concurrently use multiple stratification variables. In the second approach, concomitant variables help form strata. If these variables exhibit a positive correlation with the simulation output, they can be employed to establish optimal strata boundaries through closed-form equations. Using pilot simulation runs, we propose methods to find the best concomitant variables (that can be nonlinear transformations of real inputs or generated data during each simulation run) and the number of strata for this purpose. However, the strata can be formed with less dependence on limited runs and less computation by leveraging population statistics or looking up exact values by studentizing variables that store aggregated information. A comparison case study on the real-world data for a wind power model calibration and some static and time-dependent simulation examples illustrates faster progress and less run-to-run solver variability. Effective stratification further reduces the solver's reliance on hyperparameters, making it more robust. 

While both approaches show similar performance in many cases, choosing one approach over the other hinges on the nature of the relationship between stratification variables and the simulation output, the significance of multiple stratification variables, the availability of information regarding the distribution of these variables, and the presence of noise in simulations. As a rule of thumb, if there is evidence for heterogeneous noise and nonlinear relationships that may take time to unravel, then binary trees may be more beneficial. On the other hand, if, from expert opinion or descriptive analysis, we can find or form a concomitant variable that is linearly dependent on the response, it may single-handedly help partition the input space to tackle heteroscedasticity at a low computational cost. Among choices for concomitant variables, choosing the real data instead of simulated data may be more effective, particularly when simulation outputs are too noisy.

The present study centers on minimizing variance, which is the goal of stratified sampling in estimation and inference. However, in optimization, variance may only need to be reduced so much to help with the progress. In other words, the cost of maximal variance reduction may waste too much of the computational budget. Therefore, future work will view stratified sampling for optimization with a different objective in mind: making better local approximations that guarantee just enough accuracy to economize budget expenditure in the early iterations. Particularly for a class of adaptive simulation optimization solvers, this road-map can lead to proven lower sample complexity that can be fundamental to the theory and application of simulation optimization solvers for digital twins~\citep{goodwin2022real,santos2022use}. Deriving closed-form equations for simultaneously utilizing multiple concomitant variables for stratification is also an unexplored area for future research. Furthermore, stratification with concomitant variables requires pre-processing to identify the appropriate function of the concomitant variable that has a linear relationship with the output. This aspect is left for future research.

\bibliographystyle{apacite}
\bibliography{refs}
\end{document}